\documentclass[pra,aps,reprint,superscriptaddress,floatfix]{revtex4-2}
\usepackage{amsmath,amssymb,bbm,bm}
\usepackage{graphicx}
\usepackage{epstopdf}
\usepackage{braket}
\usepackage{latexsym}
\usepackage{bbold}
\usepackage[normalem]{ulem} 
\usepackage[sectionbib]{bibunits}
\usepackage[usenames,dvipsnames]{color}
\usepackage{hyperref}
\usepackage{physics}
\usepackage{bm}
\usepackage{xcolor, colortbl}
\usepackage{mathtools}

\makeatletter
\def\maketitle{
\@author@finish
\title@column\titleblock@produce
\suppressfloats[t]}
\makeatother

\defaultbibliographystyle{apsrev4-2} 
\defaultbibliography{thispaper} 

\newcommand{\mytitle}{Fast quantum state preparation and bath dynamics using non-Gaussian variational ansatz and quantum optimal control}

\newcommand{\affA}{Institute for Theoretical Physics, Institute of Physics, University of Amsterdam, Science Park 904, 1098 XH Amsterdam, the Netherlands}
\newcommand{\affB}{QuSoft, Science Park 123, 1098 XG Amsterdam, the Netherlands}
\newcommand{\affC}{CWI, Science Park 904, 1098 XH Amsterdam, the Netherlands}

\begin{document}


\title{\mytitle}
\date{\today}

\author{Liam J. Bond}\affiliation{\affA}\affiliation{\affB}

\author{Arghavan Safavi-Naini}\affiliation{\affA}\affiliation{\affB}

\author{Ji\v{r}\'{i} Min\'{a}\v{r}}\affiliation{\affC}\affiliation{\affA}\affiliation{\affB}

\begin{abstract}
We combine quantum optimal control with a variational ansatz based on non-Gaussian states for fast, non-adiabatic preparation of quantum many-body states. 
We demonstrate this on the example of the spin-boson model, and use a multi-polaron ansatz to prepare near-critical ground states. For one mode, we achieve a reduction in infidelity of up to $\approx 60$ ($\approx 20$) times compared to linear (optimised local adiabatic) ramps respectively; for many modes we achieve a reduction in infidelity of up to $\approx 5$ times compared to non-adiabatic linear ramps. 
Further, we show that the typical control quantity, the leakage from the variational manifold, provides only a loose bound on the state's fidelity. Instead, in analogy to the bond dimension of matrix product states, we suggest a controlled convergence criterion based on the number of polarons.
Finally, motivated by the possibility of realizations in trapped ions, we study the dynamics of a system with bath properties going beyond the paradigm of (sub/super) Ohmic couplings. We apply the ansatz to the study of the out-of-time-order-correlator (OTOC) of the bath modes in a non-perturbative regime. The scrambling time is found to be a robust feature only weakly dependent on the details of the coupling between the bath and the spin.
\end{abstract}

\maketitle  
    
\emph{Introduction}.
The description of quantum systems out-of-equilibrium represents a notorious challenge.
In many relevant situations one has to resort to numerical approaches ranging from non-equilibrium Monte Carlo to tensor networks \cite{Makri_1995_JChemPhys, Thorwart_1998_ChemPhys, Thorwart_PRE_2000, Schmidt_2008_PRB, Orus_2014_AnnPhys, Montangero_2018_Book, White_2004_PRL, Schmitteckert_2004_PRB, Nuss_2015_PRB, Dora_2017_PRB}.
A specific class of problems consists of systems containing bosonic degrees of freedom with an (even locally) unbounded Hilbert space. To deal with such situations, various schemes have been devised, such as path integral techniques \cite{Nalbach_2010_PRB, Kast_2013_PRL, Nalbach_2013_PRB, Otterpohl_2022_PRL} or effective Hamiltonian \cite{Lee_2001_PhysLettB, Rychkov_2015_PRD} and lightcone conformal truncation \cite{Anand_2020, Chen_2022_JHEP, Delacretaz_2023_JHEP} used predominantly in high-energy physics, which aim at describing the relevant part of the (bosonic) Hilbert space by a suitable choice of truncation procedure.

Another possibility is to exploit the continuous-variable structure of the bosonic states.
Here, a novel scheme using a time-dependent variational ansatz based on \emph{non-Gaussian} states has been recently proposed \cite{shiVariationalStudyFermionic2018, hacklGeometryVariationalMethods2020} and successfully applied to the studies of systems ranging from Kondo impurity problem \cite{Ashida_2019_PRA}, central spin \cite{Ashida_2019_PRL} or spin-Holstein models \cite{Knorzer_2022_PRL} to Bose and Fermi polarons \cite{Christianen_2022_PRL, Christianen_2022_PRA, Dolgirev_2021_PRX}.

In this work we demonstrate that such ansatze constitute a natural framework for the implementation of efficient state preparation schemes using quantum optimal control \cite{Lloyd_2014_PRL, Rach_2015_PRA, vanFrank_2016_SciRep, Brouzos_2015_PRA}. 
Specifically, we implement a multipolaron ansatz \cite{beraGeneralizedMultipolaronExpansion2014, Wu_2013_JChemPhys, Zhou_2014_PRB, zhouSymmetryCriticalPhase2015, wangVariationalDynamicsSubOhmic2016, Zhao_2023_JChemPhys, Chen_2023_JChemPhys} and consider the paradigmatic spin-boson model \cite{leggettDynamicsDissipativeTwostate1987a, LeHur_2008_AnnPhys, bullaNumericalRenormalizationGroup2003, Orth_2010_PRB, Nalbach_2010_PRB, Kast_2013_PRL, Nalbach_2013_PRB, beraStabilizingSpinCoherence2014, Otterpohl_2022_PRL} including in principle arbitrary couplings beyond the (sub/super) Ohmic ones and away from perturbative regimes. 
The choice of the spin-boson model is motivated by the fact that it plays a major role in the description of impurity problems, whilst also encompassing many platforms that are currently used for quantum simulation and computing, ranging from superconducting circuits to trapped ions \cite{Peropadre_2013_PRL, Yoshihara_2017_NatPhys, FornDiaz_2017_NatPhys, Magazzu_2018_NatComm, Marcuzzi_2017_PRL, Gambetta_2020_PRL, Tamura_2020_PRA, Skannrup_2020, Mehaignerie_2023, James_2000_Book, Porras_2004_PRL, Schneider_2012_RepProgPhys, Kienzler_2015_Science, Lo_2015_Nature, Kienzler_2017_PRL}. In particular, recent realizations of the quantum Rabi-Hubbard \cite{Mei_2022_PRL} and Rabi models \cite{Lv_2018_PRX, Cai_2021_NatComm} represent an ideal testbed to experimentally probe the here-presented theoretical results.

We apply the developed machinery to study (i) the onset of chaos of the bosonic bath quantified by the OTOCs, demonstrating its robustness with respect to the spin-bath couplings and (ii) fast non-adiabatic quantum many-body state preparation, including the preparation of near-critical ground states. We also highlight the limitations of the leakage as a control parameter and consider the number of polarons instead.

~\\
\emph{The model}.
We consider the spin-boson model, where the interaction of a two-level system with a bath of $N$ harmonic oscillators is described by the Hamiltonian
\begin{align}
H = \frac{\Delta}{2} \sigma_x + \sum_{k=1}^N \epsilon_k b_k^\dagger b_k - \frac{1}{2} \sigma_z \sum_{k=1}^N g_k (b_k^\dagger + b_k).
\label{eq:H_SB}
\end{align}
Here $\Delta$ describes the tunnelling strength, $\epsilon_k$ the mode frequency and $g_k$ the interaction between the spin and $k$-th mode. 
The operators $\sigma_{x,y,z}$ are Pauli operators acting on the spin, and $b_k$ ($b_k^\dagger$) the annihilation (creation) operators of the bath modes satisfying $[b_k, b^\dag_{k'}] = \delta_{k k'}$.
The Hamiltonian (\ref{eq:H_SB}) conserves the parity $P_{\text{ex}} = e^{i \pi N_{\text{ex}} }$, where $N_{\text{ex}} = 1/2(\sigma_x+1) + \sum_k b_k^\dagger b_k$ counts the total number of excitations. 

Unless stated differently, we consider Ohmic couplings, described by $\sum_k  g_k^2 \delta(\omega - \epsilon_k) = 2 \alpha \omega_c \omega \Theta(\omega_c - \omega)$. Here $\omega>0$, $\omega_c$ is a high-frequency cut-off and $\alpha$ is a dimensionless measure of the spin-bath interaction strength. We choose a mode discretization $\epsilon_k = \omega_c k/N$, $k = 1,2,\dots, N$. Note that we do not enforce any restrictions on any of the relevant energy scales, i.e. $\Delta$, $\epsilon_k$, $g_k$ or $\omega_c$. In particular, we do not require that $\omega_c$ is the largest energy scale. Such tunability is motivated by the experimental possibilities offered by trapped ion systems, where in principle arbitrary spin-boson Hamiltonians of the form (\ref{eq:H_SB}) can be engineered \cite{supp}.
~\\
\emph{Time-dependent variational principle with non-Gaussian states}. We consider a variational state $\ket{\psi(\vec{x})}$ parametrized by a set of $M$ real variational parameters $x_\mu$, $\vec{x} = \{x_1,\ldots,x_M \}$. Using the McLachlan variational principle, the imaginary and real time evolution are governed by the equations of motion \cite{shiVariationalStudyFermionic2018,hacklGeometryVariationalMethods2020,supp}
\begin{subequations}
\label{eq:EOMs}
\begin{align}
    \dot{x}^\nu &= -(g_{\mu \nu})^{-1} \partial_\mu \epsilon(\vec{x},t), \label{eq:ImaginaryEOMS} \\
\dot{x}^\nu &= -(\omega_{\mu \nu})^{-1} \partial_\mu E(\vec{x}, t).
\label{eq:RealEOMS}
\end{align}
\end{subequations}
Here $\epsilon(\vec{x},t) = E(\vec{x},t)/\bra{\psi(\vec{x})}\ket{\psi(\vec{x})}$, $E(\vec{x},t) = \bra{\psi(\vec{x})}H(t)\ket{\psi(\vec{x})}$, $g_{\mu \nu} = 2 \Re \bra{v_\mu}\ket{v_\nu}$ and $\omega_{\mu \nu} = 2 \Im \bra{v_\mu}\ket{v_\nu}$ with $\ket{v_\mu} = \partial_\mu \ket{\psi(\vec{x})}$ the tangent vectors of the variational manifold and $\partial_\mu = \partial/\partial x^\mu$ \cite{supp}.
We use (\ref{eq:ImaginaryEOMS}), (\ref{eq:RealEOMS}) to access the ground state in the $\tau \rightarrow \infty$ limit of imaginary time, and to calculate real-time dynamics respectively. 

The crucial input to the equations of motion is a non-Gaussian state, which we choose to be a multipolaron state of the form
 \begin{align}
    \ket{\psi(\vec{x})} = \sum_{p=1}^{N_p} D_{p}^{(\uparrow)} \ket{\uparrow,0} + D_{p}^{(\downarrow)} \ket{\downarrow,0},
    \label{eq:Psi_Ansatz}
\end{align} 
where
$D^{(\uparrow,\downarrow)}_p = e^{\kappa + i \theta} {\cal D}(\vec{\alpha})$, ${\cal D}(\vec{\alpha}) = \prod_{k=1}^N {\rm exp}[\alpha_k b_k^\dag - \alpha_k^* b_k]$ is the standard displacement operator of the bosonic modes and $\kappa,\theta$ encode the respective weights. Here $\vec{\alpha}=(\alpha_1,\ldots,\alpha_N)$, and we have dropped the $p,\uparrow,\downarrow$ indices for ease of notation. We note that in the limit $N_p \rightarrow \infty$, by completeness of the $N$-mode bosonic Hilbert space, $\ket{\psi(\vec{x})}$ forms an over-complete basis, and thus is in principle capable of fully describing the state of an arbitrary spin-boson system. 

The ansatz (\ref{eq:Psi_Ansatz}) who's evolution is governed by Eqs.~(\ref{eq:EOMs}) is an example of time-dependent variational principle (TDVP). It has recently found numerous applications to the time evolution of spin systems, where it is often formulated as a tensor network with time dependent parameters \cite{Kramer_2005, Leviatan_2017, Hallam_2019_NatComm, michailidisSlowQuantumThermalization2020, Turner_2021_PRX, Serbyn_2021_NatPhys}. Typically, the quality of the ansatz's evolution is quantified by a leakage
\begin{equation}
    \Lambda(t) = || (\partial_t + iH(t)) \ket{\psi(\vec{x})}||,
\end{equation}
which measures the rate at which the ansatz wavefunction leaves the variational manifold under the action of the Hamiltonian $H(t)$. The fidelity of the ansatz with respect to the true state $\ket{\Psi(t)}$ at time $t$ can be bounded by \cite{michailidisSlowQuantumThermalization2020}
\begin{align}
{\cal F}(t) = |\bra{\Psi(t)}\ket{\psi(\vec{x})}|^2 \leq \left(1 - \frac{I(t)^2}{2}\right)^2, 
\label{eq:Fidelity}
\end{align}
where $I(t) = \int_0^t d\tau \Lambda(\tau)$ is the time-integrated leakage.

\emph{Results}. Firstly, we benchmark the performance of the ansatz (\ref{eq:Psi_Ansatz}) by considering the Hamiltonian Eq.~(\ref{eq:H_SB}) with a single mode, also known as the quantum Rabi model (QRM). In this case the Ohmic coupling reduces to $g = \sqrt{2\alpha} \epsilon$ (with $\omega_c=\epsilon$), so we use $g$ and $\alpha$ interchangeably. The QRM features a crossover from a bi- to quad-polaron state at the critical coupling strength $g_c = 2\sqrt{\epsilon^2 + \sqrt{\epsilon^4 + (g_{c0}/2)^4}}$. In the so-called thermodynamic limit $\Delta/\epsilon \rightarrow \infty$ the crossover corresponds to a quantum phase transition from a normal to a superradiant phase at $g_{c0} = \sqrt{\epsilon \Delta}$ \cite{hwangQuantumPhaseTransition2015a,yingGroundstatePhaseDiagram2015}.


In Fig.~\ref{fig:maintext_dynamics}a,b we show the order parameter $\langle \sigma_x \rangle$ of the ground state in the vicinity of the crossover, and the real-time dynamics for a quench from an initial state $\ket{\psi(t=0)} = \ket{0}\ket{+}$. We see a fast convergence to the exact diagonalization (ED) results for a moderate polaron number $N_p$. The respective fidelities (\ref{eq:Fidelity}) are then shown in Fig.~\ref{fig:maintext_dynamics}c,d. 

The dotted lines in Fig.~\ref{fig:maintext_dynamics}d show the fidelity bound from the right-hand side of Eq.~(\ref{eq:Fidelity}). The bound appears to be relatively loose, in that it overestimates the actual decay of the fidelity. Although the leakage provides at least some control over the accuracy of a given ansatz, the multipolaron state (\ref{eq:Psi_Ansatz}) has the advantage that it offers the number of polarons as a control parameter. In particular $\ket{\psi(t)} \rightarrow \ket{\Psi(t)}$ in the limit $N_p \rightarrow \infty$. As such, considering the real-time dynamics of an observable ${\cal O}$, we introduce a \emph{convergence} criterion
\begin{align}
    (\Delta {\cal O}/\bar{\cal O})^{(N_p)}(T) = \frac{\int_0^T {\rm d}t | \mathcal{O}^{(\text{max}[N_p])}(t) - \mathcal{O}^{(N_p)}(t) |}
    {|\int_0^T {\rm d}t \mathcal{O}^{(N_p)}(t)|},
    \label{eq:Convergence}
\end{align}
which quantifies the relative (time-integrated) change in the evolution of the observable with respect to the maximum considered number of polarons $N_p$. Here ${\cal O}^{(N_p)}(t) = \langle \psi(t) | O | \psi(t) \rangle$ is the expectation value obtained with $N_p$ polarons.
We note that similar convergence criteria have been discussed in Refs.~\cite{zhouSymmetryCriticalPhase2015, wangVariationalDynamicsSubOhmic2016}. When considering the ground state, we shall use instead the energy variance ${\rm var}(H)$ as the convergence with ${\rm var}(H)=0$ for $\ket{\psi}=\ket{\Psi}$ \cite{beraGeneralizedMultipolaronExpansion2014}.

With these definitions at hand, we return to the spin-boson Hamiltonian (\ref{eq:H_SB}) to plot ground state properties and real-time dynamics for $N = 10$ modes, shown in Fig.~\ref{fig:maintext_dynamics}e,f respectively. 
The inset of Fig.~\ref{fig:maintext_dynamics}e plots $\text{var}(H)$. We see rapid improvement for $N_p > 1$, with relatively worse performance near the critical point $\alpha_c \approx 5$.
This is expected because our ansatz does not include squeezing, which is a property of the ground state near the critical point \cite{yingGroundstatePhaseDiagram2015}. In the inset of Fig.~\ref{fig:maintext_dynamics}f we plot $(\Delta {\cal O}/\bar{\cal O})^{(N_p)}(T)$ for $\mathcal{O} = \sigma_x$. We find $N_p> 5$ sufficient to accurately capture real-time dynamics, with $N_p > 10$ highly accurate. 
%
\begin{figure}
    \centering

    \includegraphics[width=\linewidth]{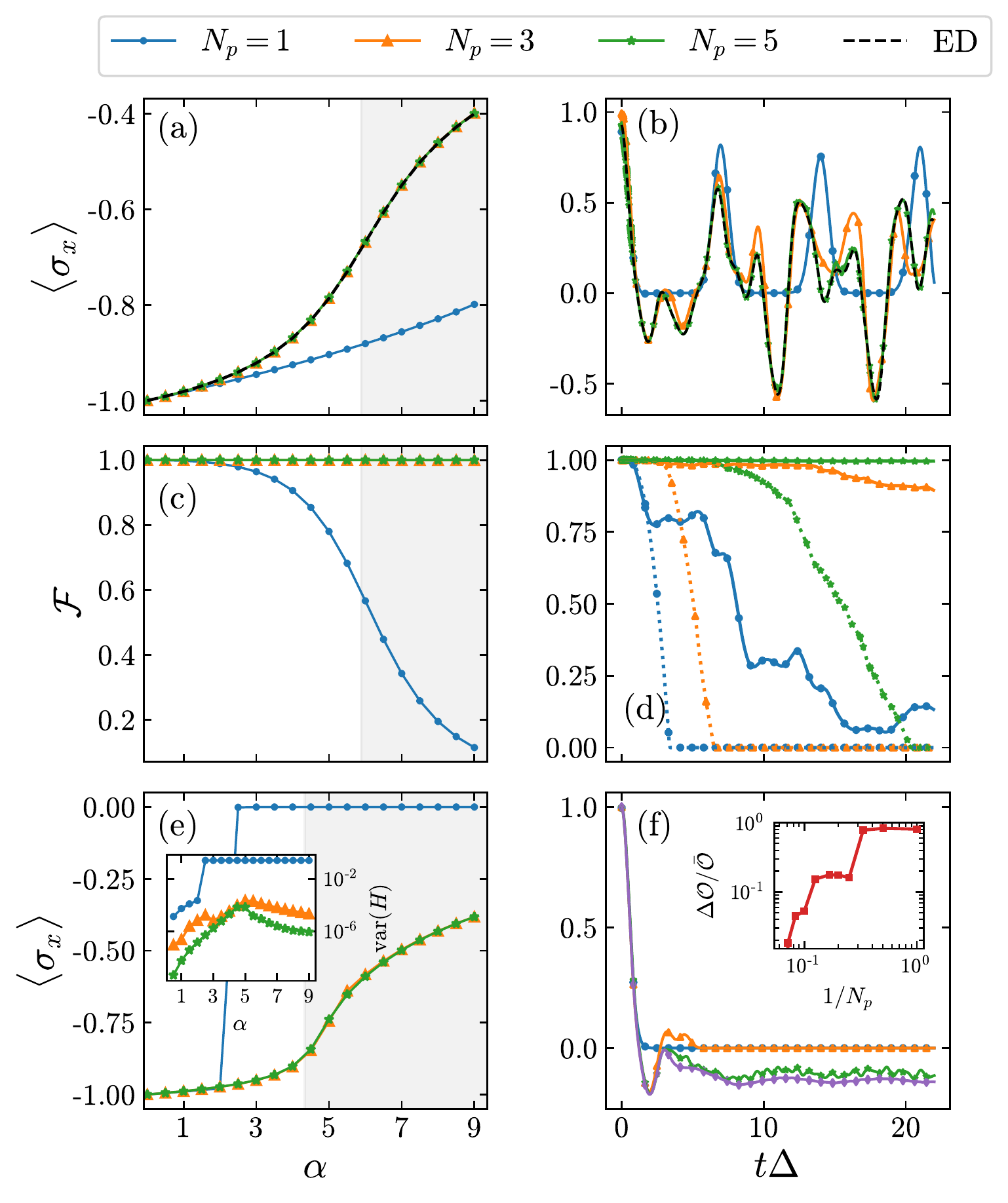}
    \caption{{\bf Single mode}: (a) Order parameter $\langle \sigma_x \rangle$ of the $P_\text{ex} = 1$ ground state and (c) fidelity $\mathcal{F} = |\bra{\Psi_{\rm gs
    }}\ket{\psi_{\rm gs}}|^2$ for $\epsilon/\Delta = 0.15/1.0$.
    The white/grey regions indicate the phase boundary between the normal/superradiant phase in the thermodynamic limit with the critical point
$g_c = \sqrt{\epsilon \Delta}$.    
    (b) Time evolution of $\langle \sigma_x \rangle$ from initial state $\ket{0}\ket{+}$ for $\epsilon/\Delta = 1.0/1.1$, $g/\Delta = 2.0/1.1$. (d) The fidelity
    $\mathcal{F} = |\bra{\Psi(t)}\ket{\psi(t)}|^2$ (solid) with the lower bound obtained from the leakage (dotted), cf. Eq.~(\ref{eq:Fidelity}). 
    {\bf Many modes}: (e) Order parameter $\langle \sigma_x\rangle$ for $N = 10$ modes. For illustration, we also include the perturbative result with critical point $\alpha_c \approx 1 + \Delta/2\omega_c$ separating the delocalized (white) and localized (grey) phases \cite{leggettDynamicsDissipativeTwostate1987a,bullaNumericalRenormalizationGroup2003}.
    The inset plots $\text{var}(H)$, which is largest near $\alpha_c$ due to the absence of squeezing in the ansatz. Panel (f) shows real-time dynamics from initial state $\ket{0}\ket{+}$, with the purple line $N_p = 16$. Parameters are $\omega_c/\Delta = 1.0/1.1$ and $\alpha = 4.0$. 
    }   
    \label{fig:maintext_dynamics}
\end{figure}

\emph{Bath dynamics.}
The ansatz Eq.~\eqref{eq:Psi_Ansatz} can be used to further quantify the bath dynamics. To this end we evaluate the fidelity OTOC, $F=\langle W^\dag(t) V^\dag W(t) V \rangle$ with $V=\ket{\psi(0)}\bra{\psi(0)}$ the projector on the initial state and $W={\rm exp}(i \delta \phi G)$ \cite{Garttner_2017_NatPhys, Garttner_2018_PRL, Lewis_2019_NatComm}. For a small perturbation $\delta \phi \ll 1$, $1-F \propto {\rm var}(G)$. We choose $G=x_k=b_k^\dag + b_k$, the position quadrature of the $k$-th mode. Such fidelity OTOCs have been considered in the analysis of chaos in the QRM in Ref. \cite{Kirkova_2022_PRA}, where it was found that the scrambling time $t^*$ corresponding to the maximum of ${\rm var}(x)$ in the superradiant phase and for a quench from a vacuum depends only weakly on the exact value of the coupling $g$ in the thermodynamic limit $\Delta/\varepsilon \rightarrow \infty$.

We demonstrate the versatility of the ansatz by moving beyond the paradigm of Ohmic-type baths. This is further motivated by the possibility to engineer arbitrary spin-bath couplings in trapped-ion systems \cite{Cai_2021_NatComm, Pedernales_2015_SciRep,supp}. We  consider a set of equally spaced $k$-modes with coupling profiles
\begin{equation}
    g_k^\pm = \frac{\bar{g}}{2} \tanh[\pm 0.25(k-\lceil N/2 \rceil)] + \bar{g},
    \label{eq:coulings}
\end{equation}
as shown in the inset of Fig.~\ref{fig:maintext_otocs}b. For both bath profiles, we study quench dynamics from $\ket{\psi(t=0)} = \ket{0}\ket{+}$ for $N=11$ modes far from the thermodynamic limit with $\Delta/\omega_c = 1.1$. In Fig.~\ref{fig:maintext_otocs}a we show the variance ${\rm var}(x_6)$ of the mid-coupling (sixth) mode with the scrambling time $t^*$ indicated. The inset shows the dependence of $t^*$ as a function of the coupling strength amplitude $\bar{g}$ (here all the couplings correspond to the (pseudo-) coherent dynamics in the phase diagram \cite{supp}) and the corresponding bosonic excitation number distribution at $t^*$ is shown in Fig.~\ref{fig:maintext_otocs}b. We find that the weak dependence of $t^*$ from the QRM in the superradiant phase and thermodynamic limit seems to be a robust feature that persists in the many mode case with very different coupling profiles and far from perturbative limits \cite{supp}. We leave this interesting opening for future systematic investigations and turn into the application of the ansatz to fast quantum state preparation. 

\begin{figure}
    \centering
   \includegraphics[width=\linewidth]{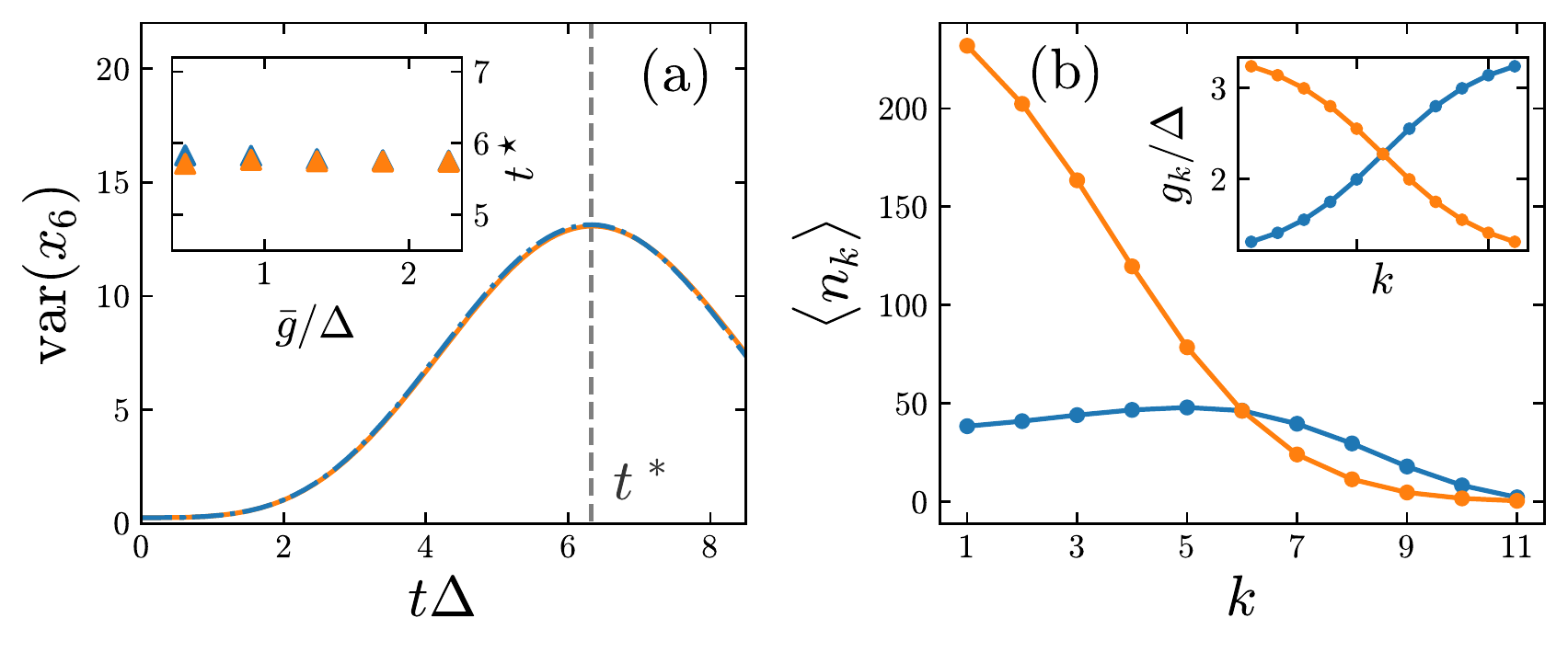}
    \caption{(a) Bath mode position quadrature of the mid-$k$ ($k=6$) mode for a quench from a state $\ket{\psi(t=0)}=\ket{0}\ket{+}$ for the coupling profiles $g_k^+$ (blue, dashed) and $g_k^-$ (orange, solid), see Eq.~(\ref{eq:coulings}) and the inset in panel (b). The vertical dashed line indicates the scrambling time $t^*$ corresponding to the first maximum of ${\rm var}(x_6)$. The inset shows $t^*$ vs. coupling amplitude $\bar{g}$ for the two profiles. (b) The occupation of the bath modes at the scrambling time.
    Parameters are $\omega_c/\Delta = 1/1.1$, $\bar{g}/\Delta=1/1.1$ and $N=11$, $N_p = 10$.
    }
    \label{fig:maintext_otocs}
\end{figure}

\emph{Quantum optimal control.} 
Adiabatic quantum state preparation, where the Hamiltonian parameters are changed such that the state during time evolution corresponds to the instantaneous ground state, is an often-employed and established paradigm with many applications in systems with global rather than local control of parameters. A prototypical example where this scheme fails is the preparation of critical states, as the adiabatic criterion cannot be satisfied due to the closure of the gap \footnote{This has motivated the design of alternative protocols such as in \cite{Agarwal_2018_PRL}.}. Going beyond adiabatic schemes requires the design of ramp protocols that generate a dynamical trajectory that takes the initial state to the target final state. 
The variational principle, which casts both the real and imaginary time evolution in the form of first order differential equations (\ref{eq:EOMs}) for the variational parameters, offers an ideal setup to implement such ramp protocols with quantum optimal control methods. 

We consider the chopped-random basis protocol (CRAB) \cite{Rach_2015_PRA}, which consists of optimizing over a set of harmonic evolutions of the control parameter. Let us first consider the quantum Rabi model. In analogy to the preparation of ground states by tuning the coupling strength $g$ \cite{Pedernales_2015_SciRep, Cai_2021_NatComm}, we consider the following time evolution of the coupling strength $g(t) = g_f t/t_f \, f(t)$ where $f(t)$ is a Fourier decomposition into $M$ harmonics,
\begin{align}
 \qquad f(t) = \frac{1}{\mathcal{N}} \left[ 1 + \sum_{j=1}^M A_j \sin(\nu_j t) + B_j \cos(\nu_j t) \right].
 \label{eq:ft}
\end{align}
Here $\mathcal{N} = 1 + \sum_j B_j$ is a normalisation factor that ensures $g(t_f) = g_f$, $\nu_j = 2 \pi j (1 + r_j)/t_f$, $r_j \in \{0,1\}$ are random integers, and the coefficients $A_j$, $B_j$ are the optimization parameters. In the above, $g_f$ is the target coupling determining the corresponding ground state.

To assess the performance of the protocol, we prepare a target ground state in the vicinity of the crossover (phase transition), which is located at coupling $g_c$. Specifically, we evaluate the  preparation time $t_f$ needed to prepare the target state with a fidelity $\mathcal{F} > 0.99$. For comparison, we also consider a linear ramp protocol $g(t)=g_f t/t_f$, and a local adiabatic (LA) ramp obtained by solving the differential equation $\gamma = |\Delta^2(g)/\dot{g}(t)|$, where $\Delta(g)$ is the instantaneous energy gap between the ground and first coupled excited state, and $\gamma \gg 1$ an adiabaticity parameter \cite{rolandQuantumSearchLocal2002,richermeExperimentalPerformanceQuantum2013a}.
In Fig.~\ref{fig:maintext_stateprep}a, we plot the infidelity as a function of the preparation time for the CRAB, linear and LA ramp protocols. The corresponding time profiles of the couplings $g(t)$ are shown in Fig.~\ref{fig:maintext_stateprep}b.
For a set $t_f$ the CRAB protocol offers a significant reduction in infidelity of $\approx 60$ times and $\approx 20$ times compared to linear and optimised adiabatic ramps respectively.
To verify that the CRAB optimization does not correspond to adiabatic evolution, in Fig.~\ref{fig:maintext_stateprep}c we show the overlap $|\langle \Psi_{\rm gs} (t) | \psi(t) \rangle|^2$ of the variational state with the instantaneous ground state $\ket{\Psi_{\rm gs}(t)}$ (we also verify that the variational state corresponds to the exact evolution $|\langle \Psi(t) | \psi(t) \rangle|^2 \approx 1$).
In Fig.~\ref{fig:maintext_stateprep}d we show the extracted preparation times for the three protocols as a function of the coupling together with the ground state boson number (grey dashed). We see that the CRAB optimization clearly outperforms both the linear and the LA ramp protocols: up to $\approx 10$ times and $\approx 2$ times faster than linear and optimised adiabatic ramps respectively.

\begin{figure}
    \centering
    \includegraphics[width=\linewidth]{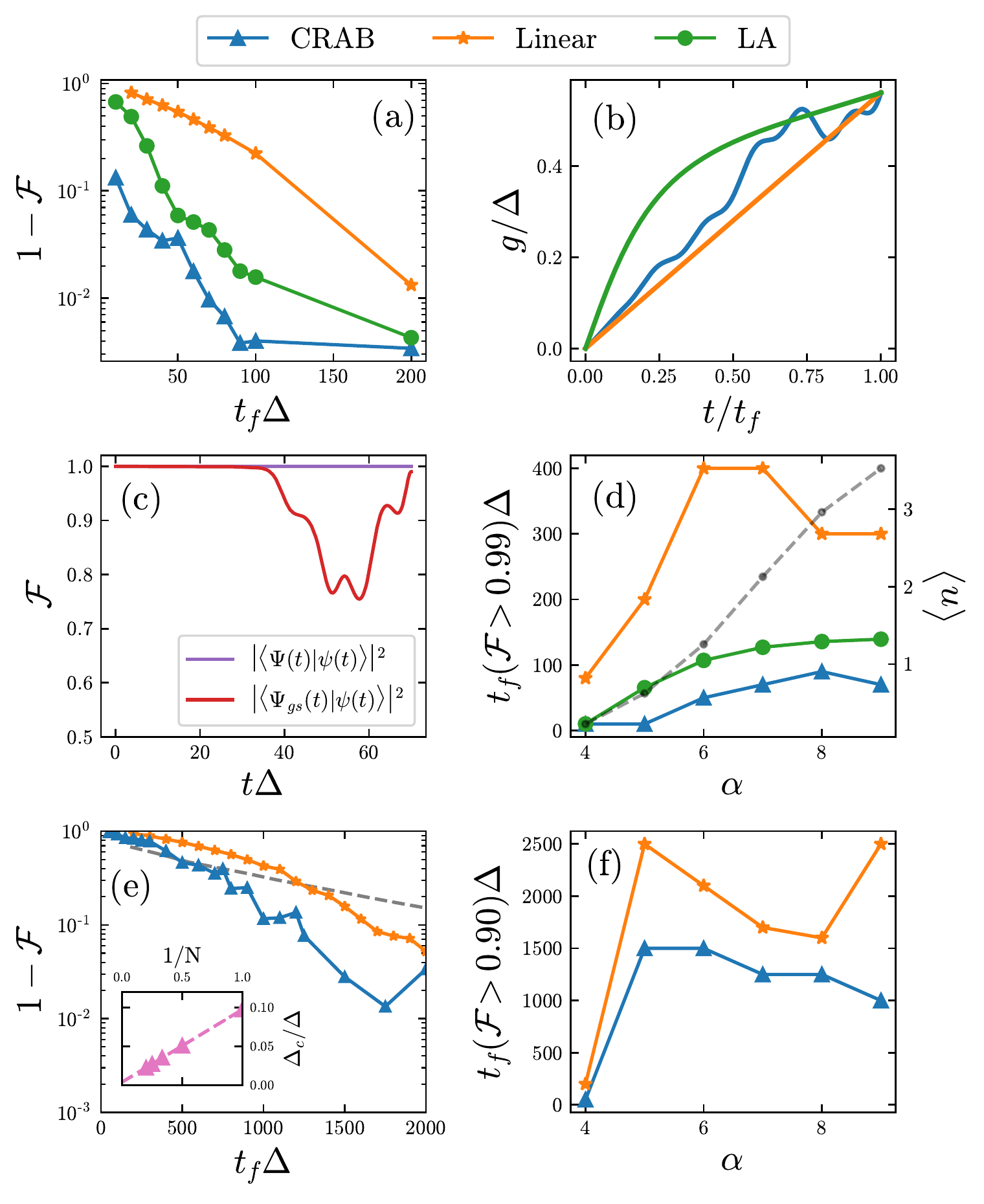}
    \caption{
    {\bf Single mode}: 
    (a) The infidelity $1-{\mathcal F}$, ${\cal F}=|\bra{\psi_{\rm gs}}\ket{\psi(t_f)}|^2 $, where $\ket{\psi_{\rm gs}}$ is the target ground state with $\alpha = 7$, and $\ket{\psi(t_f)}$ the state prepared with CRAB, Linear and LA ramps. 
    An example of the three ramps for $t_f = 70$ is shown in (b). In panel (c) we verify the accuracy of the CRAB simulation (purple) and its non-adiabaticity (red) by computing the overlap with the instantaneous ground state $\ket{\Psi_{\rm gs}(t)}$. Panel (d) shows the minimum ramp times required to prepare a target ground state at $\alpha$ with fidelity ${\cal{F}} > 0.99$. The right-axis shows the ground state boson number (grey dashed line). 
    {\bf Many modes}:
    (e) Infidelity for CRAB and Linear ramp protocols vs. ramp time for $N=10$ modes with finite size scaling of the minimum gap (inset). For comparison we show the grey dashed line obtained by extrapolating the Linear ramp data from panel (a), see \cite{supp}.
    (f) Ramp times required to prepare the target state with fidelity ${\cal F}>0.90$ vs. $\alpha$, see text for details. 
    Parameters used: $N_p = 5$, $\omega_c/\Delta = 0.15$.
    %
    } 
    \label{fig:maintext_stateprep}
\end{figure}

Moving to the many-mode case, we consider $N=10$ modes with Ohmic couplings. The target ground state for each $\alpha$ is determined using the imaginary time evolution (\ref{eq:ImaginaryEOMS}). The infidelity for a given ramp time $t_f$ for the linear and CRAB protocols is shown in Fig.~\ref{fig:maintext_stateprep}e (we omit the LA ramp for simplicity \cite{supp}). 
The inset shows the finite-size scaling of the gap \footnote{Here the gap refers to the distance to the first excited state in the same parity sector as the ground state.}. The grey dashed line, obtained by extrapolating the data for ${\cal F} = 0.9$ from Fig.~\ref{fig:maintext_stateprep}a and using the scaled gap is shown for comparison \cite{supp}. 

%
%
Next, we consider the target fidelity ${\mathcal F}>0.9$, as very high target fidelities are more stringent on the quality of the approximation (requiring sufficiently large $N_p$), cf. the Fig.~\ref{fig:maintext_dynamics}f. Fig.~\ref{fig:maintext_stateprep}f shows the preparation times $t_f$ vs. $\alpha$. Here, the adiabaticity parameter $\gamma = O(1)$ \cite{supp}, which indicates that the linear ramp times result in non-adiabatic evolution, which here is sufficient to reach the target ${\cal F}=0.9$ with only a mild improvement factor $\approx 2$ in the preparation times using the CRAB protocol \cite{supp}. This should be contrasted with $\gamma \approx 10-20$ in Fig.~\ref{fig:maintext_stateprep}d resulting in higher improvement factor of $\approx 10$ using the CRAB protocol.

\emph{Outlook.} 
We have demonstrated the application of a multipolaron non-Gaussian variational ansatz to the bath dynamics beyond (sub/super) Ohmic couplings and quantum optimal control.
As next steps, it would be interesting to investigate the bath dynamics in such non-perturbative setting including entanglement growth between the bath modes mediated by the spin or the possible absence of bound on OTOCs in such a star-graph like configuration \cite{Lucas_2019}, targeting experimental verification with trapped ions \cite{Lv_2018_PRX, Cai_2021_NatComm}. 
Another straighforward extension of our analysis is the computation of the gap through linear response \cite{shiVariationalStudyFermionic2018,hacklGeometryVariationalMethods2020} and considering carrier ramp profiles beyond the linear one, such as the LA profile in Fig.~\ref{fig:maintext_stateprep}b. This is likely to further reduce the state preparation times. 
Finally and remarkably, already the simpler Gaussian version of the ansatz \cite{Guaita_2019_PRB} allows for efficient description of systems in higher dimensions \cite{Menu_2023} or to extract scaling exponents at the phase transition \cite{Kaicher_2023}. It would be thus highly interesting to extend the here presented combination of the quantum optimal control with the multipolaron ansatz to much larger class of systems, including the open dynamics \cite{Puebla_2020_PRL}.

\begin{acknowledgments}
We would like to acknowledge stimulating discussions with D. Abanin, J.D. Bancal, E. Di Salvo, J. Home, M. Lewenstein, K. Schoutens, D. Schuricht and W. Waalewijn. This work is supported by the Dutch Research Council (NWO/OCW), as part of the Quantum Software Consortium programme (project number 024.003.037).
\end{acknowledgments}

\bibliographystyle{apsrev4-2}
\bibliography{thispaper}



\clearpage

\setcounter{equation}{0}
\setcounter{figure}{0}
\setcounter{table}{0}
\setcounter{page}{1}

\renewcommand\thefigure{\arabic{figure}}

\let\theequationWithoutS\theequation 
\renewcommand\theequation{S\theequationWithoutS}
\let\thefigureWithoutS\thefigure 
\renewcommand\thefigure{S\thefigureWithoutS}

\title{Supplemental Material: Fast quantum state preparation and bath dynamics using non-Gaussian variational ansatz and quantum optimal control}

\maketitle


\onecolumngrid

\section{Non-Adiabaticity of Linear Ramps}

In this section we study the linear ramps of Fig.~\ref{fig:maintext_stateprep} in more detail. We begin with a single mode ($N = 1$). In Fig.~\ref{fig:maintext_stateprep}d, the minimum ramp time $t_f$ required to prepare the target ground state $\ket{\psi_{\text{gs}}(\alpha)}$ with $1-\mathcal{F}<0.01$ increases with $\alpha$, before decreasing again at $\alpha \gtrsim 7$. This is counter-intuitive, because the ramp speed is set by the critical gap, which is smallest at the critical point $\alpha_c$ \footnote{Here and in the following and with a slight abuse of nomenclature we refer to critical gap and critical coupling strength to the minimal value of the gap and the corresponding coupling strength even in the crossover regime, i.e. away from the critical point in the thermodynamic sense. Such situation occurs for instance in the quantum Rabi model for finite $\Delta/\epsilon$. See also Sec.~\ref{sec:Scaling}.}. 
Provided $\alpha > \alpha_c$, preparing ground states of increasing $\alpha$ should therefore always require a larger $t_f$.  

\begin{figure}[ht]
    \centering
    \includegraphics[width=\linewidth]{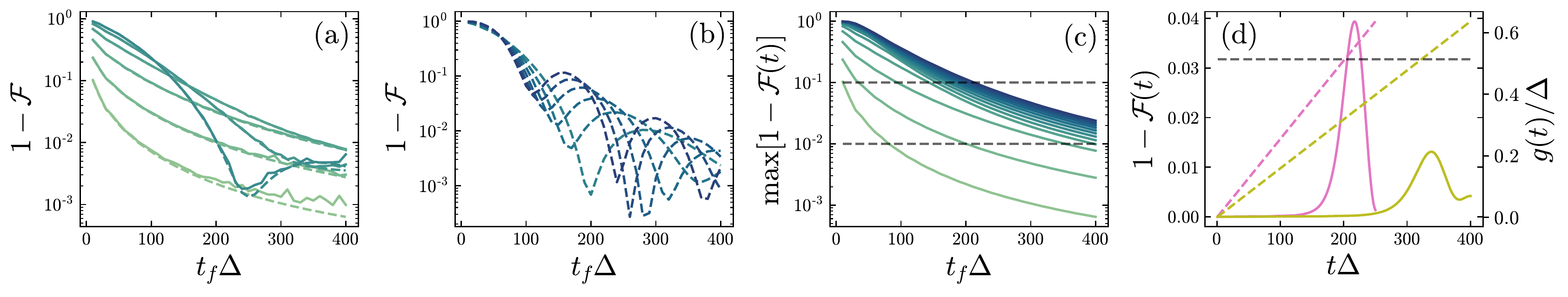}
    \caption{\textbf{Single mode:} The final infidelity $1-\mathcal{F} = 1 - |\bra{\psi_\text{{gs}}(\alpha)}\ket{\psi(t_f)}|^2$ with $\alpha = 4,5,\dots,9$ (a) and $\alpha = 10,11,\dots,15$ (b), lines colored light to dark. The dynamics are computed with ED (dashed) and NGS (solid) with $N_p = 5$ polarons. We verify that longer ramps are always more adiabatic by plotting (c) the maximum infidelity between the instantaneous state and the instantaneous ground state. Finally, we show (d) two example ramp profiles $t_f \Delta = 250$ and $t_f \Delta = 400$. Surprisingly, despite being less adiabatic, the final state of the $t_f \Delta = 250$ ramp has a higher fidelity with the target state than the $t_f \Delta = 400$ ramp. Parameters used are $\omega_c/\Delta = 0.15$}
    \label{fig:supp:LinearRamps}
\end{figure}

In Fig.~\ref{fig:supp:LinearRamps}a,b we plot the infidelity $1-\mathcal{F} = 1 - |\bra{\psi_{\text{gs}}(\alpha)}\ket{\psi(t_f)}|^2$, with $\alpha = 4,5,\dots,9$ and $\alpha = 10,11,\dots,15$ respectively, lines colored light to dark. The non-Gaussian state (NGS) calculation (solid lines) agrees well with exact diagonalization (ED) (dashed lines). For $\alpha \geq 7$, we see a surprisingly rapid and oscillatory decay in infidelity, with the number of local minima increasing for increasing $\alpha$. 

To further investigate this behaviour, we verify our intuition that longer ramps should monotonously correspond to more adiabatic evolution. To do so, in Fig.~\ref{fig:supp:LinearRamps}c we plot the maximum infidelity of the instantaneous state with the instantaneous ground state, $\max[1-\mathcal{F}(t)] = \max_{0\leq t \leq t_f}[1-|\bra{\psi_{\text{gs}}(t)}\ket{\psi(t)}|^2]$. As expected the infidelity always decreases as $t_f$ increases, while increasing as $\alpha$ increases. 

Finally, in Fig.~\ref{fig:supp:LinearRamps}d we compare $t_f \Delta = 250$ (pink) and $t_f \Delta = 400$ (olive) ramp profiles. The left axis (solid lines) shows $1-\mathcal{F}(t)=1-|\bra{\psi_{\text{gs}}(t)}\ket{\psi(t)}|^2$, while the right axis (dashed lines) shows the ramp profile $g(t)/\Delta$. The dashed horizontal line shows the critical point $g_c = 2\sqrt{\epsilon^2 + \sqrt{\epsilon^4 + (g_{c0}/2)^4}}$ where $g_{c0} = \sqrt{\epsilon \Delta}$. Note that $g = \sqrt{2\alpha} \epsilon$ and $\epsilon = \omega_c$ for $N = 1$.  We see the non-adiabaticity is located near the critical point, as expected. Despite the fact that the $t_f \Delta= 250$ ramp is less adiabatic than the $t_f \Delta = 400$ ramp, the dynamics are such that $\ket{\Psi(t)}$ returns to the ground state with higher fidelity. The minimum linear ramp time required to achieve $F \geq 0.99$ is therefore not actually realized by a fully adiabatic ramp. 

\begin{figure}[t]
    \centering
    \includegraphics[width=0.4\linewidth]{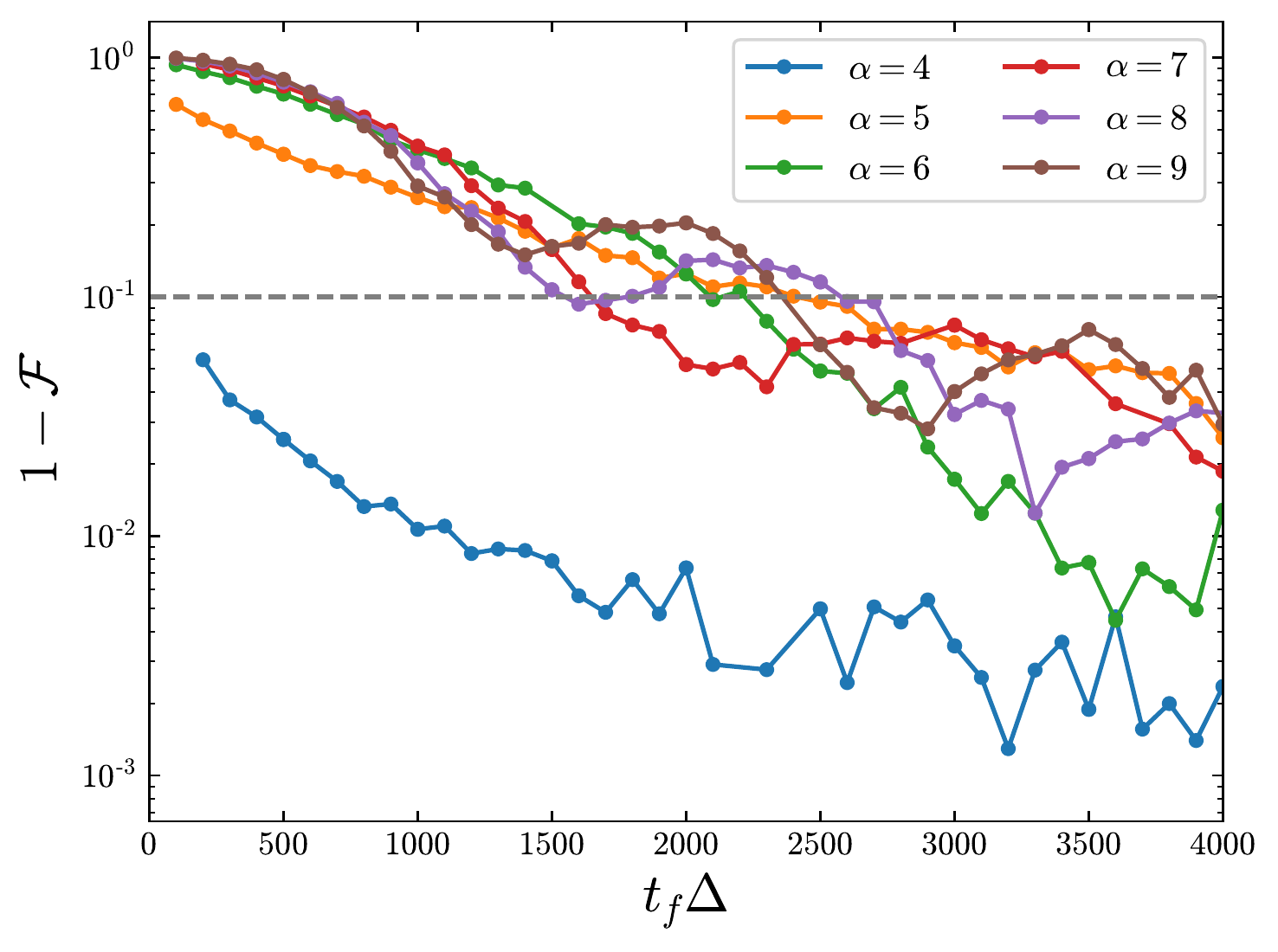}
    \caption{\textbf{Many modes:} Infidelity $1 - \mathcal{F} = 1 - |\bra{\psi_{\text{gs}}(\alpha)}\ket{\psi(t_f)}|^2$ for $N =10$ modes, calculated from real-time equations of motion with the NGS ansatz and $N_p = 5$ polarons. Note the non-monotonous decay in infidelity, similar to the single-mode result. The minimum $t_f$ required to prepare the target state with infidelity $1-\mathcal{F} < 0.1$ corresponds to the first time each $\alpha$ line crosses the dashed horizontal line. Parameters used are $\omega_c/\Delta = 0.15$.
    }
    \label{fig:supp:N10LinRamp}
\end{figure}

Next, we consider ground state preparation in the many-mode case ($N = 10$). Fig.~\ref{fig:supp:N10LinRamp} plots the final infidelity $1 - \mathcal{F} = 1 - |\bra{\psi_{\text{gs}}(\alpha)}\ket{\psi(t_f)}|^2$. The target ground state is obtained using the imaginary-time equations of motion (\ref{eq:EOMs}a) while the real-time dynamics are computed using the real-time equations of motion (\ref{eq:EOMs}b), both with $N_p = 5$. We note that this data remains consistent with the limiting case $(1-\mathcal{F}) \rightarrow 0$ as $t_f \rightarrow \infty$. We observe non-monotonous decay of the infidelity as $t_f$ increases, which is reminiscent of the single-mode behaviour. Thus, the many-mode linear ramp is also non-adiabatic. 

We have shown that both the single and many-mode linear ramp protocols are in fact non-adiabatic. As a consequence, the comparison presented in Fig.~\ref{fig:maintext_stateprep} between CRAB and linear ramp protocols is not a comparison between CRAB and adiabatic ramp protocols. If we were to restrict the linear ramp to being adiabatic, the factor by which the CRAB protocol outperforms the linear ramp would increase, particularly at large $\alpha$. 

\section{Scaling Analysis of Linear Adiabatic Ramp Times}
\label{sec:Scaling}

In this section we use the adiabatic theorem to estimate the scaling of the adiabatic linear ramp times with $\alpha$ and the number of modes. We firstly review the adiabatic theorem, obtaining a lower bound on adiabatic ramp times. Secondly, we numerically verify this bound for a single mode by changing $\alpha$. We then extend our analysis to the many-mode case to make predictions about the ramp time in regimes where ED is not tractable.

We begin with the adiabatic theorem. Following Ref.~\cite{PhysRevA.65.042308}, we denote the eigenstates of a time-dependent Hamiltonian $H(t)$ as $\ket{E_k(t),t}$, with corresponding eigenvalues $E_k(t)$, and where $k = 0$ labels the ground state. The \emph{critical gap} is defined as the minimum gap between the two smallest magnitude connected eigenvalues $E_0$ and $E_j$,
\begin{align}
    \Delta_c = \text{min}_{0 \leq t \leq t_f} [E_j(t) - E_0(t)]. \label{eq:supp:critgap}
\end{align}
From the adiabatic theorem, if we prepare the system at $t = 0$ in the ground state $\ket{E_0,0}$ and let it evolve under $H(t)$ until $t = t_f$, then the overlap between $\ket{E_0,0}$ and the final state $\ket{\psi(t_f)}$ is lower bounded by $\abs{ \bra{E_0,t_f}\ket{\psi(t_f)} }^2 \geq 1 - \epsilon^2$ if 
\begin{align}
    \left|\left \langle dH/dt \right \rangle_{j,0}\right|\Delta_c^{-2} \leq \epsilon, \forall \; t
\end{align}
where $\epsilon$ is a small number and $\langle dH/dt \rangle_{j,0} =  \bra{E_j,t} dH/dt \ket{E_0,t} $ is the matrix element describing the coupling strength between the two eigenstates. We consider the spin-boson Hamiltonian (\ref{eq:H_SB}). The ground state parity is $P_\text{ex} = 1$, while the first (second) excited state parity is $P_{\text{ex}} = -1$ $(+1)$. The relevant gap is therefore between $E_0$ and $E_2$. 

For a single mode ($N = 1$), the time-dependent parameter is the coupling strength $g(t)$ which follows a linear ramp profile $g(t) = g_f t / t_f$. The matrix element is $|dH/dt| = g_f/(2t_f) \sigma_z (b^\dagger +b)$, and thus the linear ramp is adiabatic if
\begin{align}
    t_f \gg \frac{g_f}{2\Delta_c^2} |\langle \sigma_z (b^\dagger + b)\rangle_{2,0}|. \label{eq:supp:smadiabatic}
\end{align}

For many-modes, the linear ramp profile is $g_k(t) = g_{k,f} t / t_f$ and thus the matrix element is $|dH/dt| = 1/(2t_f) \sum_k g_{k,f} \sigma_z (b_k^\dagger + b_k)$. The linear ramp is therefore adiabatic if 
\begin{align}
        t_f \gg \frac{1}{2 \Delta_c^2} \left|\left \langle \sum_k g_{k,f} \sigma_z (b_k^\dagger + b_k) \right \rangle_{2,0}\right|. \label{eq:supp:manymode_tf_adiabatic}
\end{align}

\begin{figure}[t]
    \centering
    \includegraphics[width=0.66\linewidth]{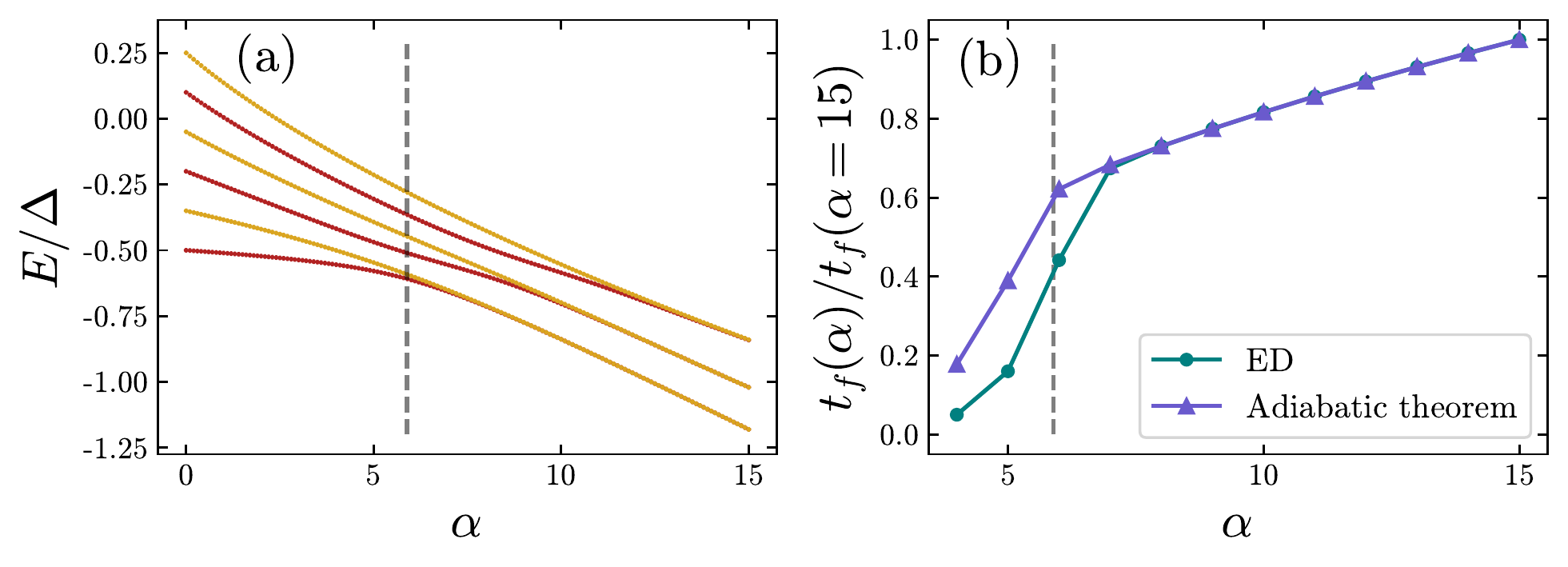}
    \caption{\textbf{Single mode:} (a) Energy spectrum with parities $P_\text{ex} = 1$ (red) and $P_\text{ex} = -1$ (gold). The dotted vertical line denotes the critical point $\alpha_c$. In panel (b) we extract $t_f$ from ED (green circles), as described in the text. In the same panel, we compare to the estimate obtained from the adiabatic theorem (purple triangles). We find the adiabatic theorem estimate to be accurate in the superradiant phase, but not the normal phase. Parameters used are $\omega_c/\Delta = 0.15$. 
    }
    \label{fig:supp:smramptimescaling}
\end{figure}

We now turn to results, beginning with $N =1$. Fig.~\ref{fig:supp:smramptimescaling}a plots the energy spectrum as a function of $\alpha$. Note that in the normal phase $\alpha \leq \alpha_c$, $\Delta_c$ as defined in Eq.~(\ref{eq:supp:critgap}) is always defined at $t_f$ as the gap narrows with increasing $\alpha$.  Fig.~\ref{fig:supp:smramptimescaling}b shows the minimum $t_f$ required to adiabatically prepare the target state $\ket{\psi_{\text{gs}}(\alpha)}$. We calculate $t_f$ using two methods. Firstly, by numerically calculating the gap and matrix element to determine the right-hand side of Eq.~(\ref{eq:supp:smadiabatic}). Because the right-hand side of Eq.~(\ref{eq:supp:smadiabatic}) is related to $t_f$ by an inequality, we normalize to the $\alpha = 15$ result, yielding a scaling of $t_f(\alpha)/t_f(\alpha=15)$ (purple triangles). The second method is a real-time numerical simulation of the linear ramp, which enables us to find the minimum ramp time $t_f$ such that $1-\bra{ \Psi_{\text{gs}}(t) }\ket{\Psi(t)} \leq 0.10 \; \forall \; 0\leq t \leq t_f$ (green circles). 
Here, the specific choice of the infidelity value at each time step is arbitrary and should be chosen $\ll 1$ to ensure the adiabaticity.
Note that this corresponds to the first time each $\alpha$ line crosses the dashed horizontal line at $1-\mathcal{F}(t) = 0.1$ in Fig.~\ref{fig:supp:LinearRamps}c.
Comparing the adiabatic to ED, we find that both agree well in the superradiant phase. However, in the normal phase the adiabatic theorem overestimates $t_f$. 

\begin{figure}[b]
    \centering
    \includegraphics[width=0.94\linewidth]{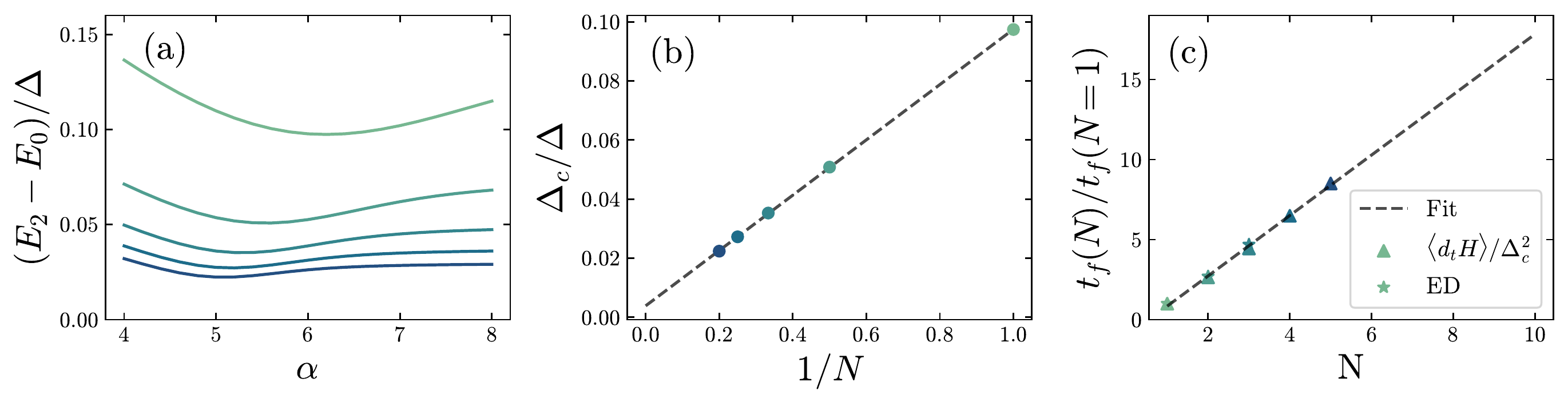}
    \caption{\textbf{Many modes}: The gap (a) between the ground and first parity-connected eigenstates for $N = 1,2,\dots,5$ (light to dark). In panel (b) we plot $\Delta_c$ as a function of $1/N$, with the black dashed line a linear fit. In panel (c) we plot Eq.~(\ref{eq:supp:manymode_tf_adiabatic}) for $N = 1,\hdots,5$ (triangles), where the gap and matrix element are calculated using ED. We perform a linear fit (black dashed line) to extrapolate $t_f$ to larger $t$. For small $N$ we simulate the real-time ramp dynamics and determine the minimum $t_f$ such that $1-\mathcal{F}(t) = 1 - |\bra{\psi_\text{gs}(t)}\ket{\psi(t)}|^2\leq0.1 \; \forall \; t$ (stars). Parameters used are $\omega_c/\Delta = 0.15$.
    }
    \label{fig:supp:ModeScaling}
\end{figure}

Next, we predict the scaling of $t_f$ with mode number. In Fig.~\ref{fig:supp:ModeScaling}a we show the gap as a function of $\alpha$ for $N=1,\hdots,5$ obtained with ED. In Fig.~\ref{fig:supp:ModeScaling}b we plot the critical gap $\Delta_c$ as a function of $1/N$, as well as a linear fit (dashed black line) allowing for extrapolation for $N$ beyond the reach of ED.
Note that in principle the many-mode spectra can also be obtained using NGS \cite{hacklGeometryVariationalMethods2020}, which we leave for future work.  

In Fig.~\ref{fig:supp:ModeScaling}c we numerically calculate the right-hand side of Eq.~(\ref{eq:supp:manymode_tf_adiabatic}) for $N=1,\hdots,5$ with $\alpha =7$ (triangles). To extrapolate beyond $N =5$, we perform a linear fit (black dashed line). We are able to verify the $t_f$ scaling for small mode numbers ($N = 1,2,3$) using a real-time ED simulation to determine the minimum ramp time such that $1-\bra{ \Psi_{\text{gs}}(t) }\ket{\Psi(t)} \leq 0.10 \; \forall \; 0\leq t \leq t_f$ when preparing $\ket{\Psi_\text{gs}(\alpha=7)}$ (stars). We see excellent agreement between the ED result and the adiabatic criterion Eq.~(\ref{eq:supp:manymode_tf_adiabatic}).

Considering a specific example of $N = 10$, from Fig.~\ref{fig:supp:ModeScaling}c we have $t_f(N=10)/t_f(N=1) \approx 18$. From Fig.~\ref{fig:supp:LinearRamps}, preparing $\ket{\psi_\text{gs}(\alpha=7)}$ with infidelity $1-\mathcal{F}<0.1$ requires $t_f(N=1) \approx 170 $. Therefore, $t_f(N=10) \approx 3\times 10^3$ for $N = 10$. In contrast, in Fig.~\ref{fig:maintext_stateprep} we find that the CRAB protocol prepares the same state with the same infidelity in $t_f = 1.25\times10^3$, which is over twice as fast. 

We note that although we have only performed this analysis for $\ket{\psi_\text{gs}(\alpha=7)}$, we expect the findings to be robust in the localised phase (ie. $\alpha > \alpha_c$). Further work is needed to investigate the scaling in the delocalised phase due to the inaccurate prediction in the normal phase of the QRM, see Fig.~\ref{fig:supp:smramptimescaling}.

\section{Adiabaticity of Linear Ramps in Fig.~\ref{fig:maintext_stateprep}}
\begin{figure}[t]
    \centering
    \includegraphics[width=0.8\linewidth]{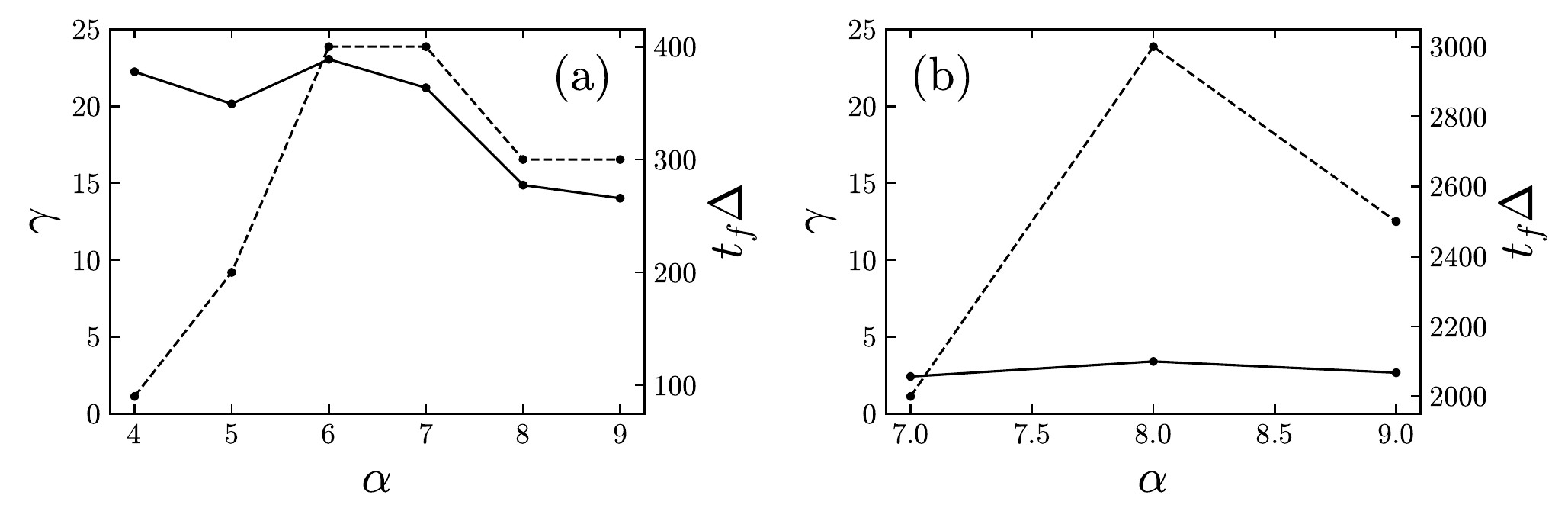}
    \caption{Adiabaticity parameter $\gamma$ (solid line, left axis) and ramp time $t_f \Delta$ (dotted line, right axis). For $N = 1$ (a), note the decrease in $\gamma$ at large $\alpha$. The $t_f$ is obtained from Fig.~\ref{fig:maintext_stateprep}(d), ie. by determining the minimum $t_f$ such that $1-\mathcal{F} = 1- |\bra{\psi_{\text{gs}}} \ket{\psi(t_f)}|^2\leq0.01$. For $N = 10$ (b), $\gamma \sim \mathcal{O}(1)$. The ramp is less adiabatic than the single-mode case because we determine $t_f$ with a higher infidelity threshold as in Fig.~\ref{fig:maintext_stateprep}(f), namely by finding the minimum $t_f$ such that $1-\mathcal{F} = |\bra{\psi_{\text{gs}}}\ket{\psi(t_f)}|^2\leq0.1$.
    }
    \label{fig:supp:gamma}
\end{figure}

In this section we evaluate the adiabaticity of the linear ramps used in Fig.~\ref{fig:maintext_stateprep}. We consider the adiabaticity parameter 
\begin{align}
    \gamma = \left | \frac{\Delta_c^2}{\langle dH/dt \rangle_{2,0} }  \right |,
\end{align}
which measures the extent to which a change in the Hamiltonian $H(t)$ is adiabatic; ie. a ramp is adiabatic if $\gamma \gg 1$. In Fig.~\ref{fig:supp:gamma} we plot the adiabaticity parameter $\gamma$ (left axis, solid lines) for both single (a) and many modes (b) using a given $t_f$ (right axis, dashed line). The $t_f$ is from Fig.~\ref{fig:maintext_stateprep}d,f, noting that the target infidelity $1-\bra{\psi_\text{gs}}\ket{\psi(t_f)}$ used to obtain the $t_f$ is $1-\mathcal{F} \leq 0.01$ for $N = 1$ and $1-\mathcal{F}\leq 0.1$ for $N = 10$. 

Here, the single-mode adiabaticity parameter $\gamma \approx 20$ indicates that the ramp is relatively adiabatic, although less so at large $\alpha$. The CRAB protocol, which is not constrained to be adiabatic, is able to prepare the same state about $\approx 10$ times faster. In contrast, for $N = 10$, $\gamma \approx O(1)$, indicating that, although sufficient to satisfy $1-\mathcal{F} < 0.1$, the linear ramps of Fig.~\ref{fig:maintext_stateprep}f are non-adiabatic. This small $\gamma$ contributes to the fact that the CRAB protocol produces the same state only $\approx 2$ times faster, a mild improvement. 

\section{On Spin Dynamics and Phases with Bath Profiles Eq.~(\ref{eq:coulings})}
Here we briefly comment on the spin dynamics and the possible underlying phases in spin-boson models with the bath couplings Eq.~(\ref{eq:coulings}).
First we note that the phase diagram of the spin-boson model has been extensively studied for the case of sub-Ohmic, Ohmic and super-Ohmic baths characterized by the spectral density $J(\omega) \propto \alpha \omega^s$ with $0<s<1$, $s=1$ and $s>1$ respectively \cite{leggettDynamicsDissipativeTwostate1987a}. There, one finds delocalized and localized equilibrium phases in the $s-\alpha$ plane as has been demonstrated in a number of works \cite{leggettDynamicsDissipativeTwostate1987a,bullaNumericalRenormalizationGroup2003,LeHur_2008_AnnPhys,Nalbach_2010_PRB,wangVariationalDynamicsSubOhmic2016} with the ground state expectation value of the magnetization, $\langle \sigma_z \rangle_{\rm gs}$, as the order parameter. Alternatively, a standard approach is to characterize the system through its non-equilibrium behaviour as quantified by the dynamics of the magnetization $\langle \sigma_z(t) \rangle$ when quenched from a fully polarized $\langle \sigma_z(t=0) \rangle = \pm 1$ state. The typical cases are a coherent (underdamped) or incoherent (overdamped) oscillations with $\langle \sigma_z(t) \rangle$ reaching the equilibrium value in the $t \rightarrow \infty$ limit. Additionally, a situation with a single oscillation before reaching the equilibrium is sometimes referred to as pseudo-coherent \cite{Otterpohl_2022_PRL}. 
It should be kept in mind that the coherent (incoherent) evolution does not in general correspond to the underlying delocalized (localized) equilibrium phases \cite{Nalbach_2010_PRB, wangVariationalDynamicsSubOhmic2016}.

\begin{figure}[ht]
    \centering
    \includegraphics[width=0.62\linewidth]{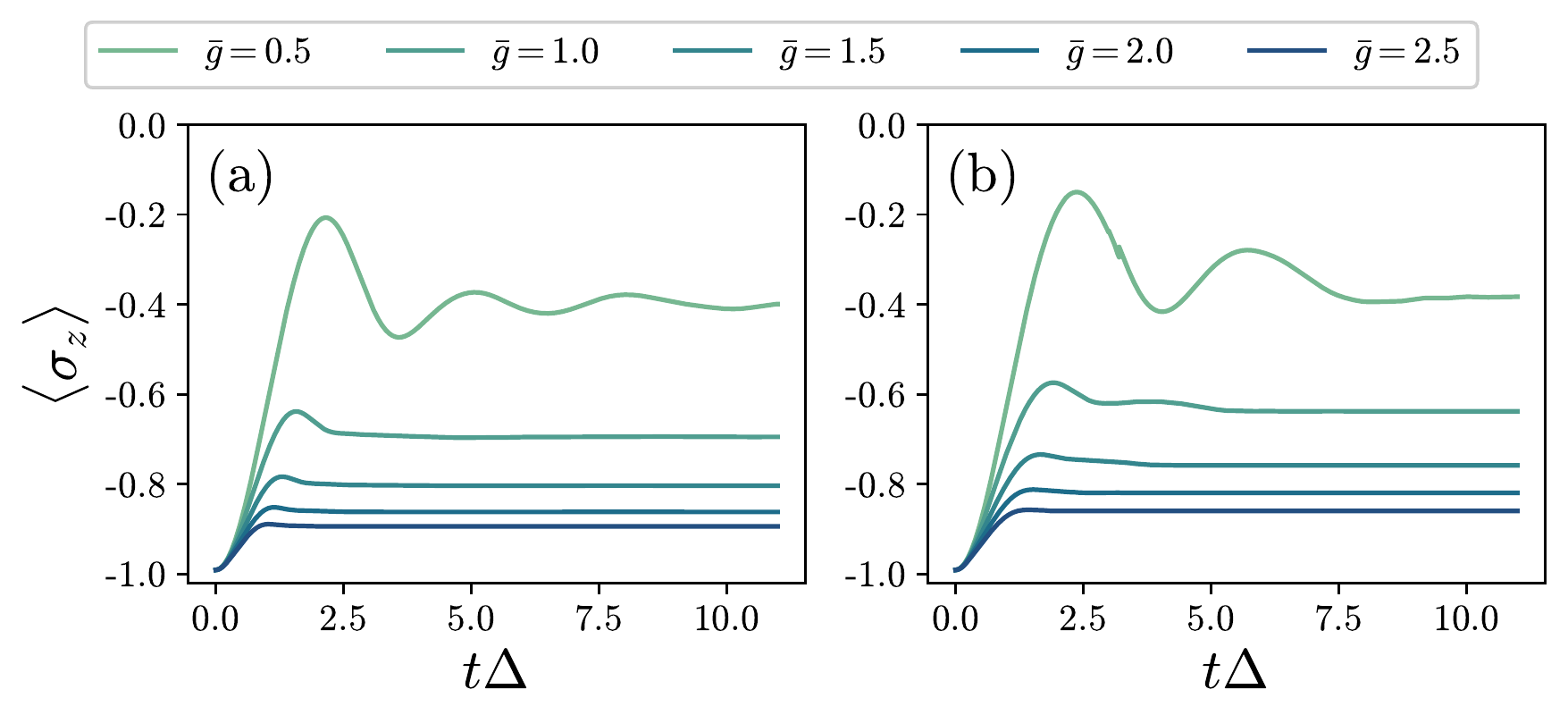}
    \caption{Magnetization dynamics $\langle \sigma_z(t) \rangle$ for a quench from an initial state with $\langle \sigma_z(t=0) \rangle=-1$ for the couplings of the form (a) $g_k^+$ and (b) $g_k^-$, see Eq.~(\ref{eq:coulings}). For increasing coupling strengths $\bar{g}$ one can see a transition from a coherent (oscillations with multiple maxima) to pseudo-coherent (oscillation with a single maximum) behaviour for both $g_k^+$ and $g_k^-$. We note that this is in qualitative agreement with the behaviour observed in spin-boson models with (sub/super) Ohmic couplings \cite{wangVariationalDynamicsSubOhmic2016, Otterpohl_2022_PRL}.}
    \label{fig:supp:bathmagnetisation}
\end{figure}

In Fig.~\ref{fig:supp:bathmagnetisation} we show the time evolution of $\langle \sigma^z \rangle$ for varying strength of the coupling $\bar{g}$ for the two profiles $g_k^+$ (Fig.~\ref{fig:supp:bathmagnetisation}a) and $g_k^-$ (Fig.~\ref{fig:supp:bathmagnetisation}b), see Eq.~(\ref{eq:coulings}). Similarly to the (sub/super) Ohmic cases \cite{wangVariationalDynamicsSubOhmic2016, Otterpohl_2022_PRL}, one can see a transition from a coherent to pseudo-coherent dynamics as $\bar{g}$ is increased from $\bar{g}=0.5$ to $\bar{g}=2.5$, the values used in the analysis of the scrambling time $t^*$ in Fig.~\ref{fig:maintext_otocs}a. There, $t^*$ is only weakly dependent on the actual value of $\bar{g}$. We could verify that this holds also in the case of the Rabi model in both the normal and superradiant phases, see also \cite{Kirkova_2022_PRA}, with the reservation that $t^*$ also depends on the initial state from which the quench is being performed.
We leave the detailed investigation of these issues, including the dynamics and the phase diagram for non-standard couplings such as in Eq.~(\ref{eq:coulings}) for future studies.


\section{Realization of Spin-Boson Model in Trapped Ions}
In trapped ion systems, the bosonic modes are collective phonon modes that arise due to the mutual Coulomb repulsion between the ions which are confined by a trapping potential. Spin-boson coupling is typically achieved using a spin-dependent force, which is realized either via a spatially dependent AC Stark shift \cite{brittonEngineered2DIsing2012,wallBosonmediatedQuantumSpin2017}, or via the simultaneous driving of a two-photon Raman transition near the red and blue sidebands \cite{leibfriedQuantumDynamicsSingle2003}. In particular, following recent theoretical proposals \cite{Pedernales_2015_SciRep,pueblaProbingDynamicsSuperradiant2017}, the single-mode variant of the spin-boson model, the quantum Rabi model, was realized in trapped ion platforms, enabling the study of real-time dynamics, ground state preparation and phase transitions \cite{Lv_2018_PRX,Cai_2021_NatComm}. 

In this section we propose an experimentally feasible realization of the many-mode spin-boson model that enables broad tunability over the parameter space. Similar to \cite{Pedernales_2015_SciRep}, our implementation utilizes a pair of Raman beams inhomogeneously detuned from the red and blue sidebands. However, we employ multiple spectral components, which enables the simultaneous driving of the red and blue sidebands of multiple modes. The broad tunability of our realization unlocks the study of the spin-boson model both beyond the paradigm of (sub/super)-Ohmic couplings, as well as in the intermediate mode number regime ($N \approx 10$ modes). 

\begin{align}
    H_0 &= H_\text{qubit} + H_\text{phonon}, \\ 
    H_\text{qubit} &= \frac{\omega_0}{2} \sum_i \sigma_i^z, \\ 
    H_\text{phonon} &= \sum_k \omega_k b^\dagger_k b_k. 
\end{align}
Here $\sigma_i^\alpha$, $\alpha = \{x,y,z\}$ are Pauli operators acting on the $i$th qubit, and $b_k^\dagger$ ($b_k$) is the creation (annihilation) operator for phonon mode $k$ with frequency $\omega_k$. 

The spin and phonon degrees of freedom are coupled by a pair of Raman beams, each with multiple spectral components. In a frame rotating with $H_\text{qubit}$, the Hamiltonian describing the interaction is
\begin{align}
    H_\text{int} = \sum_{q={1,2}} \sum_j \frac{\Omega_q}{2} \left( \sigma_j^+ e^{i\sum_k \eta_{k, j, q} (b_k^\dagger + b_k) + (\omega_0 - \omega_{q,k} ) t } + \text{h.c.}\right),
\end{align}
where $q$ indexes the pair of Raman beams, $\Omega_q$ is the Rabi frequency and $\eta_{k,j,q} = k_q b_{j k} \sqrt{\hbar / (2 m \omega_k)}$ the Lamb-Dicke parameter. Here $k_q$ is the wavevector of Raman beam $k$, and $\omega_{q,k}$ the frequencies. 

We choose $\omega_{1,k}$ ($\omega_{2,k}$) to off-resonantly drive the blue (red) sideband with detunings $\delta_k^\text{b}$ ($\delta_k^\text{r}$). That is, $\omega_{1,k} = \omega_0 + \omega_k + \delta_k^\text{b}$, $\omega_{2,k} = \omega_0 - \omega_k + \delta_k^\text{r}$. Moving to the rotating frame with respect to $H_\text{phonon}$, assuming the Lamb-Dicke regime $\eta_{k,j,q} \sqrt{\langle b^\dagger_k b_k \rangle} \ll 1 $ to expand the exponentials to the lowest order in $\eta$ with $\eta_{k,j,1} \approx \eta_{k,j,2} \equiv \eta_{k,j}$, and making a rotating wave approximation, we obtain 
\begin{align}
    \bar{H}_\text{int} = \sum_j \sum_k \frac{\eta_{j,k} \Omega_{j,k}}{2}\left[ b_k^\dagger e^{i \delta_k^\text{r} t} + b_k e^{i \delta_k^\text{b} t} \right]\sigma_j^- + \text{h.c.}.
\end{align}
Next, we require that the motional modes couple only to a single qubit. This can be achieved in several ways, for example by using two ion species (one species for the ion participating in the interaction and another species for the remaining spectator ions) \cite{bermudezAssessingProgressTrappedIon2017}; or by shelving the spectator ions into a subspace that does not couple to the Raman beams  \cite{schindlerQuantumInformationProcessor2013,niggQuantumComputationsTopologically2014}.The resulting single-spin Hamiltonian is
\begin{align}
    \bar{H}_\text{int}^1 = \sum_{k} \frac{\eta_{k} \Omega_{k}}{2}\left[ b_k^\dagger e^{i \delta_k^\text{r} t} + b_k e^{i \delta_k^\text{b} t} \right] \sigma^- + \text{h.c.}.
\end{align}
Making a unitary transformation with respect to $H_1 = 1/4\sum_n (\delta_n^\text{b} + \delta_n^\text{r}) \sigma^z + \frac{1}{2} \sum_n(\delta_n^\text{b} - \delta_n^\text{r}) b_n^\dagger b_n$ yields
\begin{align}
    \check{H}_\text{int}^1 &= \sum_{k} \frac{\eta_{k} \Omega_{k}}{2} [(b_k^\dagger+b_k) e^{-it/2 \sum_{n \neq k} (\delta_n^\text{b} + \delta_n^\text{r})} ] \sigma^- + \text{h.c.} + H_1 \\ 
    &= \sum_{k} \frac{\eta_{i,k} \Omega_{i,k}}{2} [(b_k^\dagger+b_k) e^{-it\delta(N-1)/2} ] \sigma^- + \text{h.c.} + H_1,
\end{align}
where to obtain the second line we set $\delta_n^\text{b} + \delta_n^\text{r} = \delta \, \forall \, n$.
Making a final unitary transformation with respect to $H_2 = (1/4)\delta(N-1) \sigma^z$ to clear the time-dependence from the interaction term, we obtain
\begin{align}
        H_\text{SB} = - \frac{\delta}{4} \sigma^z + \frac{1}{2} \sum_k(\delta_k^b - \delta_k^r) b_k^\dagger b_k + \frac{1}{2}\sum_{k} \eta_{k} \Omega_{k} (b_k^\dagger+b_k) \sigma^x
        \label{eq:supp:H_SB}
\end{align}
After a global $-\pi/2$ spin rotation about $\sigma_y$ which maps $\sigma^x \rightarrow \sigma^z$, $\sigma^z \rightarrow - \sigma^x$, we identify $H_\text{SB}$ as Eq.~(\ref{eq:H_SB}) with 
\begin{align}
    \Delta = \frac{\delta}{4}, \qquad \epsilon_k = \frac{1}{2}(\delta_k^b - \delta_k^r), \qquad g_k = - \eta_{k} \Omega_k. 
\end{align}
The flexibility to tune $\delta_k^\text{b}$, $\delta_k^\text{r}$ and $\Omega_k$ for each mode translates in the desired (in principle arbitrary) tunability of the parameters of the resulting spin-boson Hamiltonian (\ref{eq:supp:H_SB}).


\section{Further Details on Simulations}
All simulations are performed using Julia v1.8. The equations of motion are solved using DifferentialEquations.jl \cite{rackauckas2017differentialequations}, with the optimal control performed using Optim.jl using a Nelder-Mead algorithm \cite{mogensen2018optim}. Exact diagonalisation was performed using a combination of our own implementation and QuantumOptics.jl \cite{kramer2018quantumoptics}. 

Specifically, computing the equations of motion requires computing the tangent vectors $\ket{v_\mu}$ and the overlaps $\bra{v_\mu}\ket{v_\nu}$ to construct the symplectic form $\omega_{\mu \nu}$ and metric $g_{\mu \nu}$. We obtain the tangent vectors and overlaps analytically, enabling us to construct both $\omega$ and $g$ analytically. We obtain their pseudo-inverse, used in the equations of motion (\ref{eq:EOMs}), numerically. 
Another remark is that the equations of motion (\ref{eq:EOMs}) are norm preserving. We thus choose the parameters $\kappa, \theta$ in Eq.~(\ref{eq:Psi_Ansatz}) which ensure proper normalization as well as a \emph{global} phase factor required for a correct implementation of the TDVP \cite{Yuan_2019_Quantum, hacklGeometryVariationalMethods2020}.

Note that the equations of motion are degenerate if the initial state of at least two polarons is the same. This means that at least one polaron is unnecessary, and our equations of motion are overparametrized. In this scenario, this leads to the two polarons with the same initial state evolving identically, which reduces our ansatz of $N_p$ polarons to an effective $N_p-1$ polarons. To avoid this degeneracy, we always initialise the system such that each polaron has a slightly different initial state. For the initial vacuum state $\ket{\psi} = \ket{+}\ket{0}$, we randomly initialise each parameter as $x_\mu \in [0,0.01]$. Fidelities between different initial states are then typically $\mathcal{F}>0.99$. For quantitative studies, such as the comparison between linear ramps and CRAB when $N = 10$, we use the same initial state for both the linear and CRAB ramp.

Finally, we comment on numerical instability. We observe that there are points of numerical instability, whereby the precision required to evaluate the pseudo-inverse and equations of motion exceeds the target precision of our differential equation solver. Trajectories that pass through these points can therefore be calculated, but at increased computational cost. To avoid this, we exploit the randomness of the initial state (already required to distinguish the polarons) to generate a nearly identical initial state with nearly identical evolution, but which may not pass through the exact same point of numerical instability. We find this is sufficient to deal with the majority of cases. 


\end{document}


\title{\mytitle}
\date{\today}

\author{Liam J. Bond}\affiliation{\affA}\affiliation{\affB}
\author{Arghavan Safavi-Naini}\affiliation{\affA}\affiliation{\affB}
\author{Ji\v{r}\'{i} Min\'{a}\v{r}}\affiliation{\affA}\affiliation{\affB}

\maketitle  

\appendix

\section{Non-Adiabaticity of Linear Ramps}

In this section we investigate the behaviour of the linear ramp protocol in more detail. We begin with the single-mode $N = 1$ case. In the main text, we observe that the ramp time $t_f$ required to prepare the target state decreases at large $\alpha$. This is counter-intuitive, because the speed of the ramp profile is set by the critical gap, which is smallest at the critical point $\alpha_c$. Preparing ground states at $\alpha > \alpha_c$ should therefore always require a longer ramp time. 

\begin{figure}[h!]
    \centering
    \includegraphics[width=\linewidth]{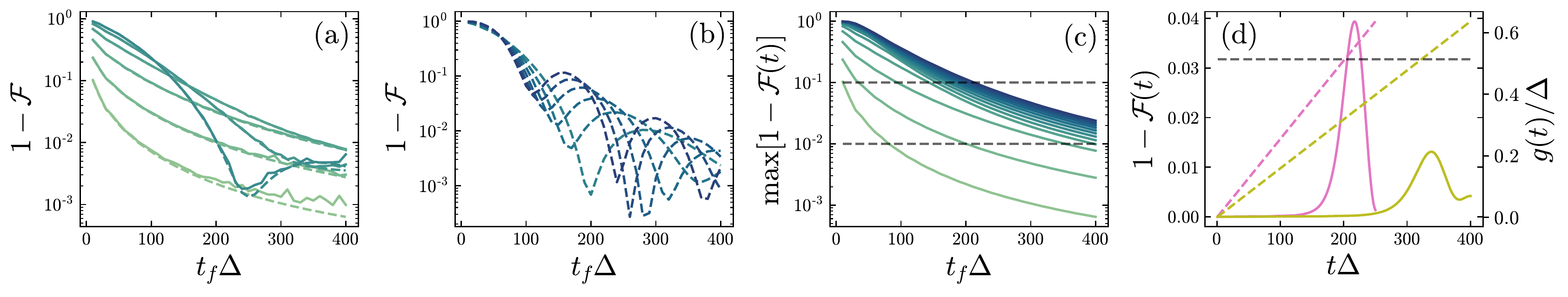}
    \caption{Non-adiabaticity of the linear ramp for $N = 1$ modes. Panels (a),(b) show the infidelity $1-\mathcal{F}$, $\mathcal{F} = |\bra{\psi_\text{{gs}}(\alpha)}\ket{\psi(t_f)}|^2$, where $\ket{\psi_{\text{gs}}(\alpha)}$ is the target ground state. Here $\alpha = 4,5,\dots,15$ (light to dark) split across the two panels, calculated with ED (dashed) and NGS with $N_p = 5$ polarons (solid). Panel (c) confirms that the maximum infidelity with the instantaneous ground state $\ket{\psi_\text{gs}(t)}$ always decreases with $t_f$, while increasing with $\alpha$. In panel (d) we plot two example ramp profiles for the target state $\ket{\psi_{\text{gs}}(\alpha)}$ prepared in $t_f \Delta = 250$ and $t_f \Delta = 400$. Surprisingly, despite being less adiabatic the final state of the $t_f \Delta = 250$ ramp has a higher fidelity with the target state than the $t_f \Delta = 400$ ramp. Parameters used are $\omega_c/\Delta = 0.15$}
    \label{fig:supp:LinearRamps}
\end{figure}

In Fig.~\ref{fig:supp:LinearRamps}(a),(b) we plot the infidelity $1-\mathcal{F} = 1 - |\bra{\psi_{\text{gs}}(\alpha)}\ket{\psi(t_f)}|^2$, with $\alpha = 4,5,\dots,9$ (panel a) and $\alpha = 10,11,\dots,15$ (panel b) colored light to dark. Note that the NGS (solid lines ) result agrees well with the ED (dashed lines). For $\alpha \geq 7$, the infidelity decreases more rapidly than expected for an adiabatic ramp. Further, in panel (b) we see an oscillatory decay, with the number of local minima increasing for increasing $\alpha$. In panel (c) we confirm our expectation that the ramp becomes more adiabatic with ramp time by plotting the maximum infidelity of the state with the instantaneous ground state, $\max(1-\mathcal{F}) = \max_{0\leq t \leq t_f}[1-\bra{\psi_{\text{gs}}(t)}\ket{\psi(t)}]$. As expected the infidelity always decreases as $t_f$ increases, while increasing as $\alpha$ increases. In panel (d) we compare the ramp profiles of $t_f \Delta = 250$ (pink) and $t_f \Delta = 400$ (olive) when preparing the target state $\ket{\psi_{\text{gs}}(\alpha=9)}$. The left axis (solid lines) shows $1-\mathcal{F}=1-|\bra{\psi_{\text{gs}}(t)}\ket{\psi(t)}|^2$, while the right axis (dashed lines) shows the ramp profile $g(t)/\Delta$. The dashed horizontal line shows the critical point $g_c$. We see the non-adiabaticity is located near the critical point, as expected. Despite the fact that the $t_f \delta= 250$ ramp is less adiabatic than the $t_f \Delta = 400$ ramp, the dynamics are such that $\ket{\Psi(t)}$ returns to the ground state with higher fidelity. The minimum linear ramp time required to achieve $F \geq 0.99$ is therefore not actually realised by a fully adiabatic ramp. 

\begin{figure}[h!]
    \centering
    \includegraphics[width=0.4\linewidth]{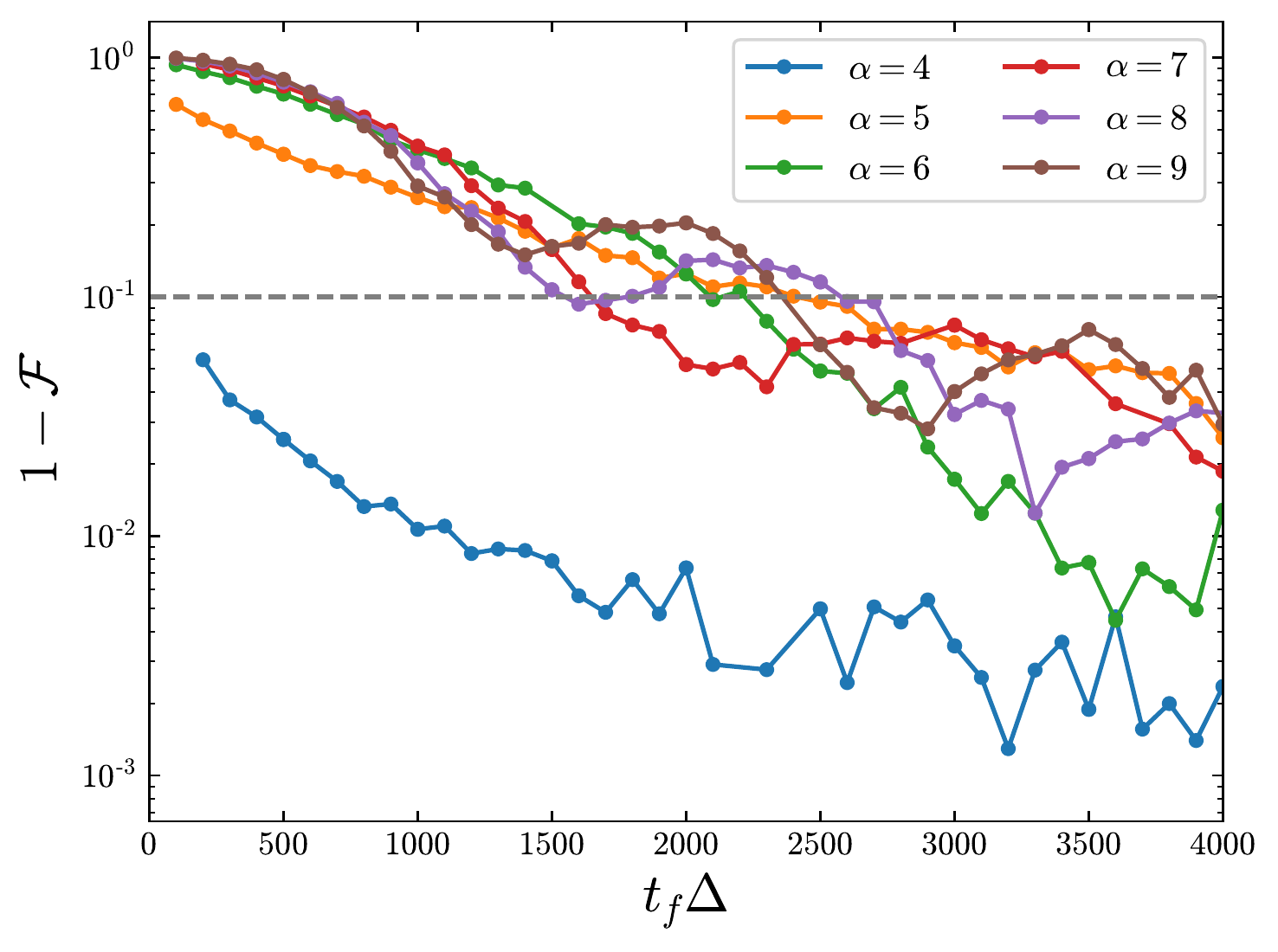}
    \caption{Infidelity $1 - \mathcal{F} = 1 - |\bra{\psi_{\text{gs}}}\ket{\psi(t_f)}|^2$ for $N =10$, computed using NGS with $N_p = 5$ polarons. Note the non-trivial decay in infidelity, which is consistent with the single-mode result. The minimum $t_f$ required to prepare the target state with infidelity $1-\mathcal{F} < 0.1$ corresponds to the first time each $\alpha$ line crosses below the dashed horizontal line. Parameters used are $\omega_c/\Delta = 0.15$.}
    \label{fig:supp:N10LinRamp}
\end{figure}

Next, we consider ground state preparation in the many-mode case ($N = 10$). In Fig.~\ref{fig:supp:N10LinRamp} we plot the infidelity $1 - \mathcal{F} = 1 - |\bra{\psi_{\text{gs}}}\ket{\psi(t_f)}|^2$. The target ground state is obtained using the imaginary-time EOMs while the real-time dynamics are computed using the real-time EOMs, both using the NGS multipolaron ansatz with $N_p = 5$. We note that this data remains consistent with limiting behaviour $(1-\mathcal{F}) \rightarrow 0$ as $t_f \rightarrow \infty$. For smaller $t_f$, we observe similar non-trivial oscillatory behavior as in the single-mode case, which is consistent with the linear ramp producing non-adiabatic dynamics. 

\section{Many-mode critical gap scaling}
In this section we use the adiabatic theorem to estimate how the ramp times of an adiabatic linear ramp scale with the number of modes. We firstly review the adiabatic theorem, obtaining a lower bound on adiabatic ramp times. We then numerically verify this bound for a single mode, before finally extending our analysis to the many-mode case using scaling analysis. 

We begin with the adiabatic theorem. Following Ref.~\cite{PhysRevA.65.042308}, we denote the eigenstates of a time-dependent Hamiltonian $H(t)$ as $\ket{E_k(t),t}$, with corresponding eigenvalues $E_k(t)$, and where $k = 0$ labels the ground state. The \emph{critical gap} is defined as the minimum gap between the two smallest magnitude connected eigenvalues $E_0$ and $E_j$,
\begin{align}
    \Delta_c = \text{min}_{0 \leq t \leq t_f} [E_j(t) - E_0(t)]. \label{eq:supp:critgap}
\end{align}
From the adiabatic theorem, if we prepare the system at $t = 0$ in the ground state $\ket{E_0,0}$ and let it evolve under $H(t)$ until $t = t_f$, then the overlap between $\ket{E_0,0}$ and the final state $\ket{\psi(t_f}$ is lower bounded by $\abs{ \bra{E_0,t_f}\ket{\psi(t_f)} }^2 \geq 1 - \epsilon^2$ if 
\begin{align}
    \left|\left \langle dH/dt \right \rangle_{j,0}\right|\Delta_c^{-2} \leq \epsilon,
\end{align}
where $\epsilon$ is a small number and $\langle dH/dt \rangle_{j,0} =  \bra{E_j,t} dH/dt \ket{E_0,t} $ is the matrix element describing the coupling strength between the two eigenstates. 

For $H(t)$ we use the spin-boson Hamiltonian, defined in the main text. The ground state has parity $P_{\text{ex}} = 1$, while the first (second) excited state has parity $P_{\text{ex}} = -1$ $(+1)$. The relevant gap is therefore between $E_0$ and $E_2$. The time-dependent parameter is the coupling strength $g(t)$. 

\begin{figure}[h!]
    \centering
    \includegraphics[width=0.66\linewidth]{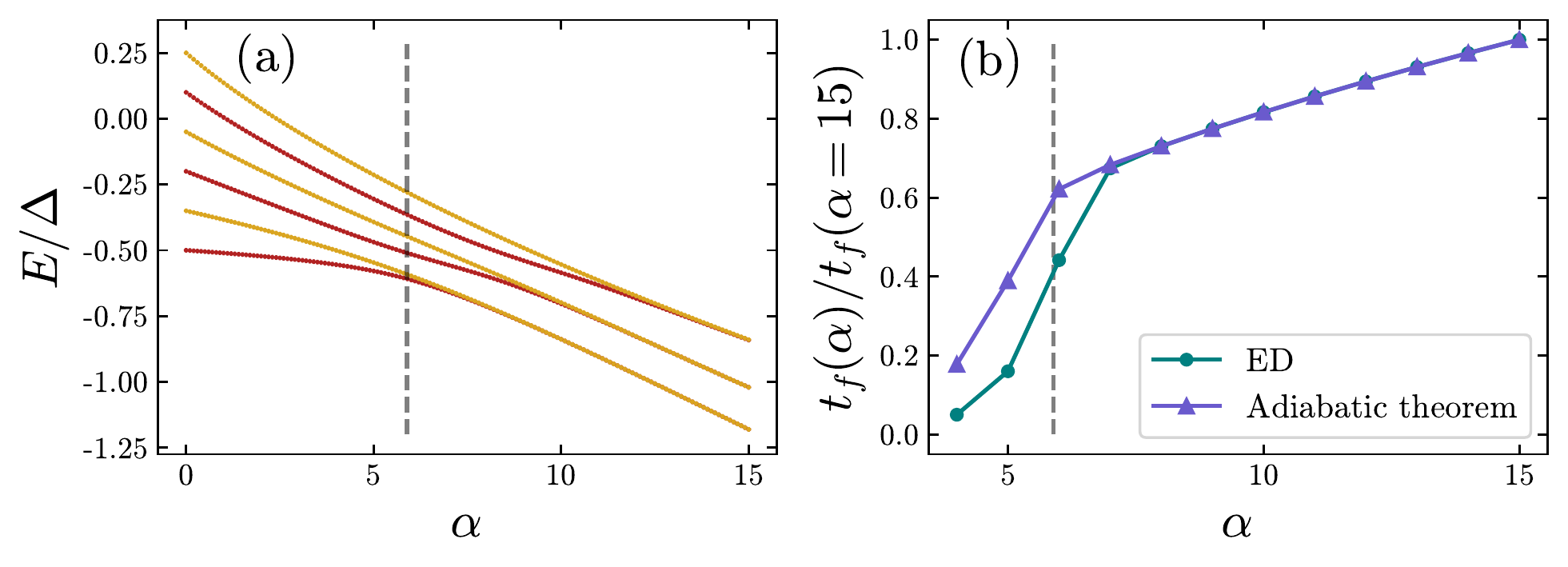}
    \caption{Single mode linear ramp time scaling. Panel (a) shows the spectrum, with parities $P_\text{ex} = 1$ (red) and $P_\text{ex} = -1$ (gold). Dotted grey vertical line denotes the critical point $\alpha_c$. In panel (b) we extract $t_f$ from ED, as described in the text. In the same panel, we compare to the estimated obtained from the adiabatic theorem. We find that although the adiabatic theorem estimate is accurate in the superradiant phase, it overestimates $t_f$ in the normal phase. Parameters used are $\omega_c/\Delta = 0.15$. }
    \label{fig:supp:smramptimescaling}
\end{figure}

Focusing firstly on the single-mode $N = 1$ case, we consider a linear ramp $g(t) = t_f t / t_f$. The matrix element is then $|dH/dt| = g_f/(2t_f) \sigma_z (a^\dagger +a)$. Thus, for the ramp to be adiabatic we require that 
\begin{align}
    t_f \gg \frac{g_f}{2\Delta_c^2} |\langle \sigma_z (a^\dagger + a)\rangle_{2,0}|. \label{eq:supp:smadiabatic}
\end{align}
We verify this scaling relation numerically in Fig.~\ref{fig:supp:smramptimescaling}. In panel (a) we plot the energy spectrum as a function of $\alpha$. Note that in the normal phase, the critical gap is as defined in Eq.~\ref{eq:supp:critgap}, ie. the minimum gap where $0 \leq t \leq t_f$. In panel (b) we show the minimum ramp time $t_f$ required to prepare the target state $\ket{\psi_{\text{gs}}}$ with target fidelity $\mathcal{F} \geq 0.90$. We calculate $t_f$ using two methods: extracted from ED (blue dots), and the adiabatic theorem prediction of Eq.~\ref{eq:supp:smadiabatic} (purple triangles). To obtain $t_f$ from ED, we determine $\text{min}_{0 \leq t \leq t_f}[t]$ such that $\bra{ \Psi_{\text{gs}}(t) }\ket{\Psi(t)} \geq 0.90$ for all $0\leq t \leq t_f$. In the superradiant phase (critical point denoted by grey vertical line) the scaling prediction is accurate. However, in the normal phase the adiabatic theorem overestimates the required ramp time.

\begin{figure}[h!]
    \centering
    \includegraphics[width=\linewidth]{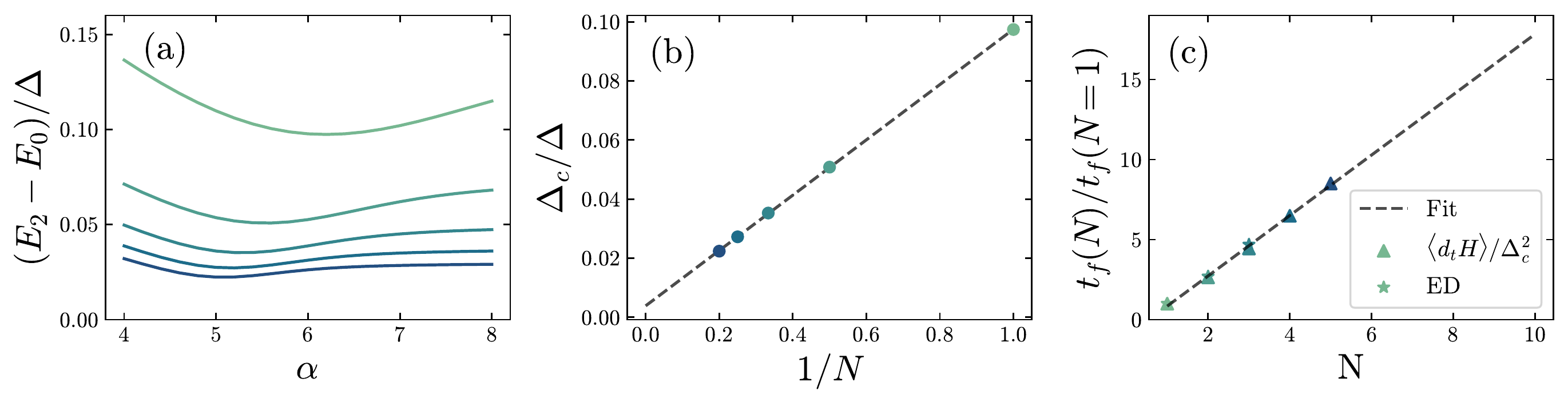}
    \caption{Many-mode gap scaling. Panel (a) shows the gap between ground and first parity-connected eigenstate for $N = 1,2,\dots,5$ (light to dark) calculated using ED. The gap closes as $N$ increases. In panel (b) we plot $\Delta_c$ as a function of $1/N$, with the black dashed line a linear fit. In panel (c) we predict the scaling of the linear ramp time from the adiabatic theorem (triangles) with a linear fit (black dashed). We verify the scaling for a small number of modes $N \leq 3$ using ED (stars), where $t_f$ is extracted from ED in the same way as the single-mode result. Parameters used are $\omega_c/\Delta = 0.15$.}
    \label{fig:supp:ModeScaling}
\end{figure}

Switching now to many-modes, we again investigate a linear ramp $g_k(t) = g_{k,f} t / t_f$. The matrix element is $|dH/dt| = 1/(2t_f) \sum_k g_{k,f} \sigma_z x_k,$, and thus 
\begin{align}
        t_f \gg \frac{1}{2 \Delta_c^2} \left|\left \langle \sum_k g_{k,f} \sigma_z (a_k^\dagger + a_k) \right \rangle_{2,0}\right|. 
\end{align}
For large $N$, extracting the gap using ED is not tractable due to the exponential scaling of the Hilbert space. Instead, we compute the critical gap $\Delta_c$ up to $N = 5$, and extrapolate to larger $N$. In Fig.~\ref{fig:supp:ModeScaling}(a) we show the gap for increasing $N$, and plot the critical gap $\Delta_c$ with a linear fit in panel (b). We perform a similar scaling analysis for the matrix element. In panel (c) we calculate the scaling of $t_f(N)/t_f(N=1)$ up to $N = 10$, finding that $t_f(N=10) \sim 18\times t_f(N=1)$. Using the data from Fig.~\ref{fig:supp:LinearRamps}, preparing the $\alpha = 7$ ground state with infidelity $1-F<0.1$ requires $t_f(N=1) \sim 170$, which therefore scales to $t_f(N=10) \sim 3\times 10^3$ for $N = 10$.  In contrast, our optimised non-adiabatic ramp prepares the same state with the same infidelity in $t_f = 1.25\times10^3$, which is twice as fast. 

\begin{figure}[h!]
    \centering
    \includegraphics[width=0.8\linewidth]{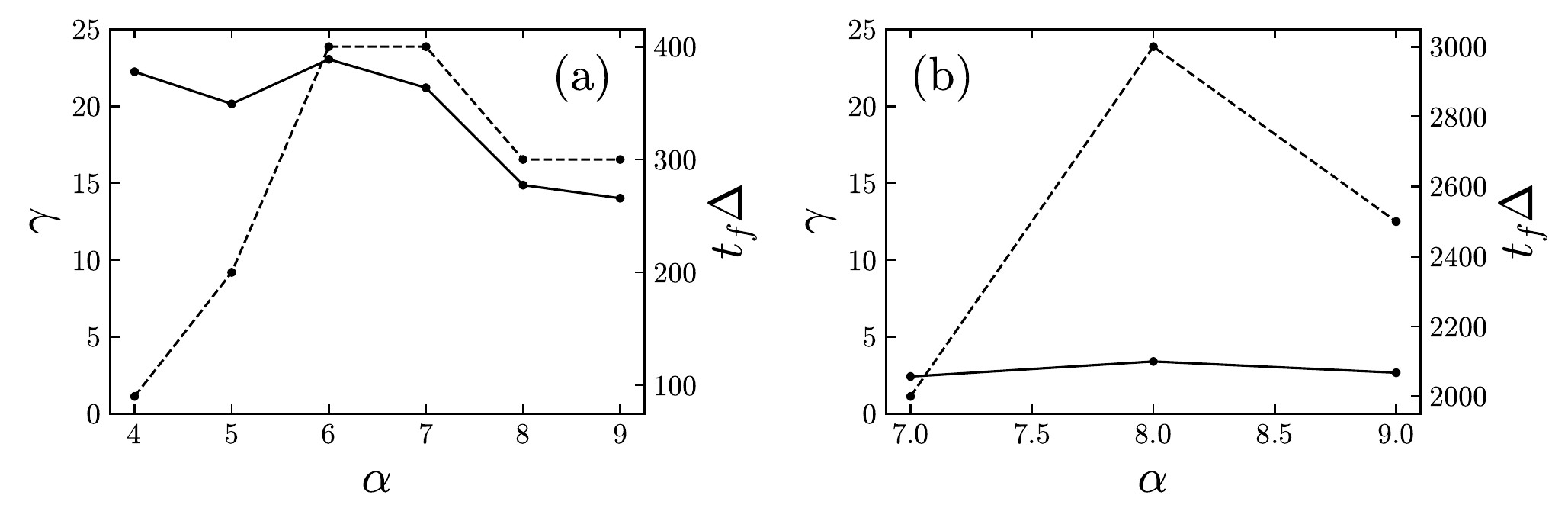}
    \caption{Adiabaticity parameter $\gamma$ (solid line, left axis) and ramp time $t_f \Delta$ (dotted line, right axis). For $N = 1$ (a), note the decrease in $\gamma$ at large $\alpha$ as the ramp is less adiabatic. The $t_f$ is such that $\mathcal{F} = |\bra{\psi_{\text{gs}}\ket{\psi(t)}}|^2\geq0.99$. For $N = 10$ $\gamma \sim 3$ because the ramp is less adiabatic than the single-mode case, predominantly due to the weaker requirement $\mathcal{F} = |\bra{\psi_{\text{gs}}}\ket{\psi(t)}|^2\geq0.90$ that was used to obtain $t_f$.}
    \label{fig:supp:gamma}
\end{figure}

Finally, in Fig.~\ref{fig:supp:gamma} we plot the adiabaticity parameter
\begin{align}
    \gamma = \left | \frac{\Delta_c^2}{\langle dH/dt \rangle_{1,0} }  \right |
\end{align}
for both the single mode (panel a) and many mode (panel b) cases. The left axis shows $\gamma$ for a given $t_f$ (right axis) which we extract from the main text. For a ramp to be adiabatic, we require that $\gamma \gg 1$. 

\section{Signatures of the Delocalized Phase in the Bath Dynamics} 
\begin{figure}[h!]
    \centering
    \includegraphics[width=0.62\linewidth]{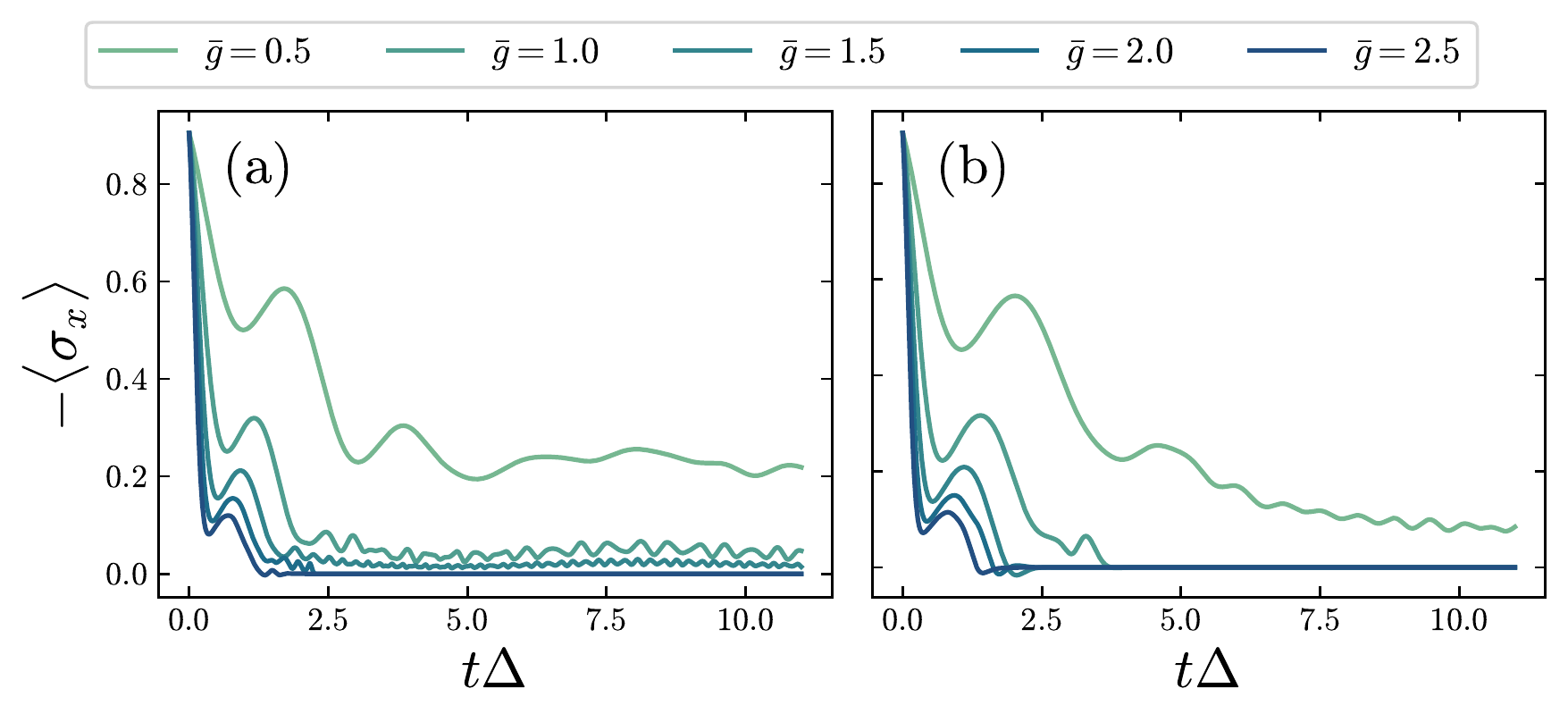}
    \caption{Magnetisation dynamics. We observe oscillations in the magnetisation, which is a signature of the delocalized phase of the many-mode spin-boson model. The coupling profile of panel (a) is $g(k)_+$, whilst panel (b) is $g(k)_-$, as defined in the main text. }
    \label{fig:supp:bathmagnetisation}
\end{figure}

In the main text, in Fig.~(2) we investigate OTOCs for various coupling profiles. We claim that, for our chosen parameters, the dynamics occurs in the delocalized phase phase. In Fig.~\ref{fig:supp:bathmagnetisation} we provide evidence for this claim by plotting the dynamics of the magnetisation $-\langle \sigma_x \rangle$. We always observe at least one oscillation, which is a signature of the delocalized phase. 

\section{Realization of Spin-Boson Model in Trapped Ions}
Spin-boson coupling terms of this form have been considered before \Com{[REF]} and realised in trapped ion experiments \Com{[REF]}. Here we explicitly detail how the spin-boson model can be realised in such a way that enables access to all areas of the phase diagram. \Com{ref asn, pedernales, experimental paper?? asn dicke one? or one where they address many sidebands? }

We consider $N$ trapped ions forming a system of spins and phonons 
\begin{align}
    H_0 = \sum_m \omega_m b^\dagger_m b_m + \frac{\omega_\text{eg}}{2} \sum_i \sigma_i^z. 
\end{align}
The operators $\sigma_i^{x,y,z}$ are Pauli operators acting on the qubit, while $a_m$ ($a_m^\dagger$) is the annihilation (creation) operator for the $m$th bosonic mode. In trapped ions, the mode frequencies ${\omega_m}$ are the phonon mode frequencies. 

A pair of Raman beams is used to realize a spin-phonon coupling of the form 
\begin{align}
    H_{SB} = \sum_k \sum_j \frac{\Omega_k}{2} \left( \sigma_j^+ e^{i[\sum_m \eta_{m, j, k} (a_m^\dagger + a_m) + (\omega_\text{eg} - \omega_k ) t] } \right),
\end{align}
where $k = 1,2$ indexes the pair of Raman beams, $\Omega_q$ is the Rabi frequency and $\eta$ is the Lamb-dicke parameter. In the Molmer-Sorensen scheme, the pair of Raman beams have the same Rabi frequencies, and address the blue and red sidebands with opposite detunings. Here, we instead detune from the blue (red) sideband by $\delta_m^b$ ($\delta_m^r$), i.e. $\omega_1 = \omega_{\text{eg}} + \omega_m + \delta_m^b$, $\omega_2 = \omega_{\text{eg}} - \omega_m + \delta_m^r$. Moving to the rotating frame of $H_0$, expanding to the lowest order in the Lamb-Dicke parameter and making a rotating wave approximation to discard terms rotating at frequencies faster than $\delta_m^{k}$, we obtain 
\begin{align}
    H_I = \sum_{i=1} \sum_{m=1} \frac{\eta_{i,m} \Omega_{i,m}}{2}\left[ a_m e^{i \delta_m^b t} + a_m^\dagger e^{i \delta_m^r t}\right]\sigma_-^i + \text{h.c.}.
\end{align}

We shelve all ions bar one into a state that does not interact with the light, yielding a single-spin Hamiltonian
\begin{align}
    H_I^1 = \sum_{m=1}^N \frac{\eta_{m} \Omega_{m}}{2}\left[ a_m e^{i \delta_m^b t} + a_m^\dagger e^{i \delta_m^r t}\right] \sigma_- + \text{h.c.}.
\end{align}
Next, we make a unitary transformation with respect to $H_0 = 1/4\sum_n (\delta_n^b + \delta_n^r) \sigma_z + \frac{1}{2} \sum_n(\delta_n^b - \delta_n^r) a_n^\dagger a_n$, which results in
\begin{align}
    \bar{H}_I^1 = \sum_{m=1}^N \frac{\eta_{m} \Omega_{m}}{2} [(a_m^\dagger+a_m) e^{-it/2 \sum_{n \neq m} (\delta_n^b + \delta_n^r)} ] \sigma_- + \text{h.c.} + H_0 
\end{align}
Next, we require that $\delta_n^b + \delta_n^r = \delta \, \forall \, n$, yielding
\begin{align}
    \sum_{m=1}^N \frac{\eta_{i,m} \Omega_{i,m}}{2} [(a_m^\dagger+a_m) e^{-it\delta(N-1)/2} ] \sigma_- + \text{h.c.} + H_0 
\end{align}
Making a final unitary transformation with respect to $H_1 = (1/4)\delta(N-1) \sigma_z$, we obtain 
\begin{align}
        - \frac{\delta}{4} \sigma_z + \frac{1}{2} \sum_m(\delta_m^b - \delta_m^r) a_m^\dagger a_m + \frac{1}{2}\sum_{m=1}^N \eta_{m} \Omega_{m} (a_m^\dagger+a_m) \sigma_x 
\end{align}
After a single-qubit rotation, we identify this as the spin-boson Hamiltonian of \Com{Eq.~[REF]} in the main text with parameters 
\begin{align}
    \Delta = \frac{\delta}{4}, \qquad \epsilon_k = \frac{1}{2}(\delta_m^b - \delta_m^r), \qquad g_m = - \eta_{m} \Omega_m. 
\end{align}
Note that the requirement that $\delta_m^b + \delta_m^r = \delta$ still gives us freedom to choose the difference in detuning $\delta_m^b - \delta_m^r$ for each mode $m$, thus giving us the ability to tune the $\epsilon_k$ spectrum. We can therefore access the full phase diagram of the spin-boson model by setting the detunings $\delta^b_k, \delta^r_k$ and Rabi frequencies $\Omega_k$. 

\section{Further Details on Simulations}
All simulations are performed using Julia v1.8. The equations of motion are solved using DifferentialEquations.jl \cite{rackauckas2017differentialequations}, with the optimal control performed using Optim.jl using a Nelder-Mead algorithm \cite{mogensen2018optim}. Exact diagonalisation was performed using a combination of our own implementation and QuantumOptics.jl \cite{kramer2018quantumoptics}. 

Specifically, computing the equations of motion requires computing the tangent vectors $\ket{v_\mu}$ and the overlaps $\bra{v_\mu}\ket{v_\nu}$ to construct the symplectic form $\omega_{\mu \nu}$ and metric $g_{\mu \nu}$. We obtain the tangent vectors and overlaps analytically, enabling us to construct both $g$ and $J$ analytically \Com{Should I include an example of this? Or outline our procedure?}. We obtain their pseudo-inverse numerically. 

Note that the equations of motion are degenerate if the initial state of at least two polarons is the same. This means that at least one polaron is unnecessary, and our equations of motion are overparametrized. In this scenario, this leads to the two polarons with the same initial state evolving identically, which reduces our ansatz of $N_p$ polarons to an effective $N_p-1$ polarons. To avoid this degeneracy, we always initialise the system such that each polaron has a slightly different initial state. Typically, we randomly initialise each parameter as $x_\mu \in [0,0.01]$. Fidelities between different initial states are then typically $\mathcal{F}>0.99$. For sensitive simulations such as the comparison between linear ramps and CRAB when $N = 10$, we use the same initial state for both the linear and CRAB ramp. 

Finally, we comment on numerical instability. We observe that there are points of numerical instability, whereby the precision required to evaluate the pseudo-inverse and equations of motion exceeds the target precision of our differential equation solver. Trajectories that pass through these points can therefore be calculated, but at increased computational cost. To avoid this, we exploit the randomness of the initial state (already required to distinguish the polarons) to generate a nearly identical initial state with nearly identical evolution, but which may not pass through the exact same point of numerical instability. We find this is sufficient to deal with the majority of cases. During the CRAB optimization, if we do encounter a numerically unstable evolution, we abort that iteration and set the fidelity of that ramp profile to $\mathcal{F} = 0$.

\bibliography{suppbib}